\documentclass[a4paper,fleqn,usenatbib,useAMS]{mnras}

\usepackage{graphicx}	
\usepackage{amsmath}	
\usepackage{amssymb}	
\usepackage{multicol}        
\usepackage{bm}		
\usepackage{pdflscape}	

\newcommand{\buddi}{\textsc{buddi}} 

\usepackage{ulem}
\usepackage[usenames,dvipsnames]{xcolor}

\newcommand{\paperII}[1]{\textcolor{red}{Paper~II}}
\defcitealias{Thomas_2011}{TMJ}

\usepackage[T1]{fontenc}
\usepackage{ae,aecompl}

\usepackage{newtxtext,newtxmath}

\title[BUDDI-MaNGA I]{ BUDDI-MaNGA I: A statistical sample of cleanly decomposed bulge and disc spectra}

\author[E. J. Johnston \textit{et al}]{Evelyn J. Johnston$^{1}$\thanks{Contact e-mail: \href{mailto:evelyn.johnston@mail.udp.cl}{evelyn.johnston@mail.udp.cl}}, Boris H\"au\ss ler$^{2}$, \&  Keerthana Jegatheesan$^{1}$.
\\
$^{1}$ N\'ucleo de Astronom\'ia, Facultad de Ingenier\'ia y Ciencias, Universidad Diego Portales, Av. Ej\'ercito Libertador 441, Santiago, Chile\\
$^{2}$ European Southern Observatory, Alonso de Cordova 3107, Vitacura, Santiago, Chile\\
}

\pubyear{2022}

\begin{document}
\label{firstpage}
\pagerange{\pageref{firstpage}--\pageref{lastpage}}
\maketitle

\begin{abstract}
Many galaxies display clear bulges and discs, and understanding how these components form is a vital step towards understanding how the galaxy has evolved into what we see today. The BUDDI-MaNGA project aims to study galaxy evolution and morphological transformations through the star-formation histories of the bulges and discs. We have applied our \textsc{buddi} software to galaxies from the MaNGA Survey in the SDSS DR15 in order to isolate their bulge and disc spectra, from which we derived their stellar populations. To date, this work provides the largest sample of clean bulge and disc spectra extracted from IFU datacubes using the galaxies light profile information, and will form the basis for a series of papers aiming to answer open questions on how galaxies have formed and evolved, and the role of their individual structures. This paper presents an introduction to the project, including an overview of these fits, a characterisation of the sample, and a series of tests on the fits to ensure reliability.
\end{abstract}




\section{Introduction}
In the hierarchical model of galaxy evolution, galaxies have built up their mass through mergers and accretion of other galaxies \citep[e.g.][]{Guo_2008}. Many galaxies display multiple components, such as bulges and discs, and as the galaxy grows in mass,  interactions and star formation activity within the galaxy affects these components differently. Therefore, understanding how these structures assemble their mass is a key element in understanding galaxy evolution in general. Many galaxies are also known to consist of only one component, such as a pure disc or a pure bulge (i.e. an elliptical galaxy), and so investigating the transformation between one and two-component systems (in both directions) will also help us explain the existence of these systems.

Many studies of bulges and discs have been carried out via bulge--disc decomposition techniques on photometric data, in which the light profile of each component is fitted either in one or two dimensions and used to create a best-fitting model for each component. From these models, the structural parameters, such as the size and luminosity of each component, can be derived. In the last two decades, this technique has been employed by many large photometric surveys of galaxies, such as SDSS \citep[e.g.][]{Simard_2011}, GAMA \citep[e.g.][]{Lange_2016}, CANDELS \citep[e.g.][]{Dimauro_2018}, and CANDELS+HFF \textcolor{black}{(Nedkova et al, in prep)}, in order to obtain statistically significant results on their properties over a range of redshifts. Applying this technique to multi-waveband data provides magnitudes for the bulge and disc in each band. These colours can in turn be used to derive their stellar populations through SED fitting \citep[e.g.][]{Kennedy_2016}.

However, deriving estimates of the ages and metallicities of stellar populations from such broad-band colours alone can result in highly degenerate stellar populations \citep{Worthey_1994}, and can also be affected by dust reddening \citep{Disney_1989}. The addition of spectroscopic information is vital towards breaking this degeneracy and measuring the star-formation histories across galaxies. In recent years, several approaches have been used to cleanly separate the light from different components within galaxies observed with long-slit spectroscopy, using either the differences in their light profiles \citep[e.g.][]{Silchenko_2012, Johnston_2012, Johnston_2014} or their distinct kinematics \citep[e.g.][]{Coccato_2011, Johnston_2013, Tabor_2017}. With the introduction of integral-field spectrographs with high spatial resolution and large fields of view, such as the Multi-Unit Spectroscopic Explorer \citep[MUSE,][]{Bacon_2010}, and surveys like the Calar Alto Legacy Integral Field spectroscopy Area survey \citep[CALIFA,][]{Sanchez_2012}, the Sydney-AAO Multi-object Integral-field spectrograph survey \citep[SAMI,][]{Croom_2012}, and the SDSS-IV Mapping Nearby Galaxies at Apache Point Observatory survey \citep[MaNGA,][]{Bundy_2015}, these spectroscopic decomposition techniques can now be adapted to make the most of the combined spatial and spectroscopic information that is now available. 

For example, several works have focussed on the different kinematics of the bulges and discs to separate their spectra.  Typically, one expects the galaxy to consist of a spherical, dispersion-supported bulge surrounded by a rotating disc, and so one can try to model the kinematics of each component using this assumption to extract the spectra of the kinematic bulge and disc. This idea has been used successfully by \citet{Tabor_2017} and \citet{Tabor_2019} for CALIFA and MaNGA galaxies, respectively. They used photometric decomposition to determine the ratio of light from the bulge and disc throughout the galaxy, and used these light ratios as constraints when modelling the spectra with 2 kinematic components. A similar approach has also been used to study the bulge and disc of NGC~3521 by \citet{Coccato_2018} and  a large sample of galaxies from the SAMI survey by \citet{Oh_2020}. And another study by \citet{Du_2020} applied this technique to galaxies in the TNG100 run of IllustrisTNG to compare the properties of the morphological and kinematic components present in disc galaxies.

Other approaches use the 2-dimensional light profile of the galaxy to separate the bulge and disc spectra. For example, \citet{Fraser_2018b} used decomposition of SDSS photometric data to determine which spaxels lay in bulge- and disc-dominated regions within MaNGA galaxies in order to minimize contamination from other structures. They then concentrated their analysis on these areas to derive the properties of the stellar populations of the disc and bulge regions.  A similar approach was used by \citet{Pak_2021}, while \citet{Barsanti_2021} combined this idea with two other approaches (i.e. using photometric decomposition of the galaxy to determine the mass and flux fractions of the bulge and disc across the IFU field of view (FOV) in addition to simply identifying bulge- and disc-dominated regions) for their spectroscopic decomposition of SAMI galaxies.

The next natural step is to apply this bulge--disc decomposition directly to the IFU data itself to make full use of the spatial and spectroscopic information available in the datacubes.  One approach that uses this idea is Bulge--Disc Decomposition of IFU data \citep[\textsc{buddi},][]{Johnston_2017}, which uses \textsc{galfitm} \citep{Haeussler_2013}, a modified form of \textsc{galfit} \citep{Peng_2002, Peng_2010} that can model the light profile of multi-waveband images of a galaxy simultaneously. \textsc{buddi} uses information from the entire datacube to create a wavelength-dependent model of each component within the galaxy, from which the spectra of each component can be cleanly extracted. It was developed and tested on prototype data from the MaNGA Survey in \citet{Johnston_2017}, and has since been successfully used to separate the light from many different components within  galaxies, such as bulges, discs and lenses of S0 galaxies \citep{Johnston_2021}, to extended stellar haloes around cD galaxies \citep{Johnston_2018}, and down to the nuclear star clusters in the cores of dwarf galaxies \citep{Johnston_2020}. 

A similar technique to \textsc{buddi} is \textsc{c2d}, which was developed by \citet{Mendez_2019a}. In this case, the initial fit to the galaxy light profile is carried out on complementary imaging data, as opposed to the white-light image of the galaxy itself as in the case of \textsc{buddi}, but it uses the same general idea of modelling the galaxies light profile as a function of wavelength. \textsc{c2d} has been successfully applied to galaxies in the CALIFA survey to study the properties of ETGs \citep{Mendez_2019b}, and of bulges and discs over a wide range of galaxy morphologies \citep{Mendez_2021}.

We are now entering an exciting time, where these different IFU decomposition techniques are transitioning from proof-of-concept ideas to being used for science. While they have so far mainly been applied to small samples of galaxies, where each galaxy could be modelled carefully by the user, the field is now moving towards automated fits to large data sets to carry out statistical studies of the spectroscopic stellar populations of galaxy components. To this end, this work presents the BUDDI-MaNGA project, in which we apply \textsc{buddi} to all suitable galaxies included in the recent MaNGA data release 15 \citep[hereafter DR15,][]{Aguado_2019}.  We identify the `good' fits to the data  (see Section~\ref{sec:success}), and extract clean bulge and disc spectra for these galaxies and estimates their stellar populations and star-formation histories. This work will produce the largest sample of clean bulge and disc spectra to date, allowing statistical studies of the stellar populations of each component for the first time, and will be expanded in the future with SDSS Data Release 17 \citep{Abdurrouf_2021}.

This paper is the first in the series, and provides an overview of the project, while \textcolor{black}{Johnston et al (submitted)} presents the first scientific analysis of the S0 galaxies in the sample. Future works will further explore galaxy formation, evolution, and morphological transformations through the effects on the different components within the galaxies.

This paper is laid out as follows: Section~\ref{sec:data} outlines the data and catalogs used, Section~\ref{sec:method} gives an overview of \textsc{buddi} and details of the fits that are specific to the MaNGA data. Section~\ref{sec:overview} presents the results for the structural parameters derived using \textsc{buddi} and the criteria used to determine which fits are considered to be good enough for further analysis, and the stellar populations analysis is given in Section~\ref{sec:overview_stellar_pops}. We then present a comparison of the stellar populations estimates when the fits are repeated using the structural parameters derived from fits to photometric data instead in Section~\ref{sec:BUD_PYM_comparison}, and finally we present our conclusions in Section~\ref{sec:conclusions}.

\section{Data}\label{sec:data}
The data used in this paper came from the SDSS DR15 of the  MaNGA Survey. It should be noted that the MaNGA data products included as part of data release 16 \citep[DR16,][]{Ahumada_2020} are identical to those in DR15, and this data counts as the third data release for MaNGA. Data from the full survey became public in SDSS DR17, and will be used for future works within the scope of the BUDDI-MaNGA project. 

MaNGA is an integral field spectroscopic survey using the BOSS spectrograph \citep{Smee_2013} on the 2.5 m SDSS telescope \citep{Gunn_2006} at the Apache Point Observatory, and is part of the Sloan Digital Sky Survey-IV \citep[SDSS-IV,][]{Blanton_2017}. By the end of the survey in 2020, MaNGA observed $\sim10,000$ galaxies, of which 4824 have been released publicly as part of DR15. The galaxies were selected to cover a wide range in mass and morphology, with redshifts in the range $0.01 < z < 0.15$ \citep{Yan_2016}. The observations for each galaxy provide spatial coverage out to $\sim1.5R_e$ for $\sim66$~per cent of the total sample, and out to $\sim2.5R_e$ for the remaining galaxies. This coverage is achieved using a series of hexagonal IFUs of different sizes, from the 19-fibre IFUs which have a 12\arcsec\ diameter up to the 127-fibre IFUs with a diameter of 32\arcsec\ \citep{Drory_2015}. The final data has a continuous wavelength coverage between 3600 to 10300~\AA, with a spectral resolution of $R\sim1400$ at 4000~\AA\ to $R\sim2600$ at 9000~\AA\ \citep{Drory_2015}. 

The decomposition technique used in this paper is the \textsc{buddi} code \citep{Johnston_2017}, which was originally developed  and tested on commissioning data observed as part of the MaNGA survey. As part of these tests, a series of simulated datacubes were created based on the MaNGA datacubes for galaxies observed with the 127, 91 and 61-fibre IFUs, and these datacubes were modelled with \textsc{buddi} to compare the extracted parameters with those used to create the models. We found that \textsc{buddi} was able to successfully extract the physical parameters and line strengths for 2 component fits in the 127 and 91-fibre IFUs, while the scatter in the results increased in the smaller IFUs due to the small number of resolution elements, critical to a successful decomposition.  In particular, the fits to the 61-fibre IFUs were found to result in overestimates for the bulge S\'ersic index and $R_e$ in about half the simulated datacubes, and underestimates for the line strengths measured from the extracted disc spectra. This effect on the line strengths is likely due to the overmodeling of the bulges due to the effect on the S\'ersic indices and sizes, resulting in poor fits to the discs and low S/N for their spectra. As a result, the data sample selected for this study was restricted to only those galaxies observed with the 91 and 127-fibre IFUs.  This limitation consequently resulted in a total sample of 1,926 candidate galaxies in SDSS DR15, which we used as our starting sample. The breakdown of the numbers of galaxies used in the fits is given in Table~\ref{tab:numbers}.

\begin{table}
	\centering
	\caption{Overview of the galaxy sample modelled with the single S\'ersic (SS) and S\'ersic+exponential (SE) models}
	\label{tab:numbers}
	\begin{tabular}{lrr} 
 		\hline
						 		& SS		& SE \\
		\hline
  Number of galaxies (DR15)	 			& 4,824	& 4,824 \\
  Number of candidate galaxies (IFU only)		& 1,926 	& 1,926\\
  Number of candidate galaxies (MPP-VAC) 	& 1,719 	& 1,705\\
  Number of failed/crashed fits  				& 402 	& 405\\
  Number of bad fits 						& 279	& 609\\
  Number of successful fits 				& 1,038	& 691\\
 		\hline
\multicolumn{3}{p{3.2in}}{The number of candidate galaxies (IFU only) refers only to the number of galaxies observed with the 91 and 127-fibre IFUs, while the MPP-VAC candidate galaxies give the numbers of galaxies that have been flagged as having successful fits in the MPP-VAC for each model. The failed and bad fits refer to those that were attempted using the MPP-VAC parameters as initial estimates, but in which \textsc{Galfitm} crashed during the fitting process (i.e. the fit failed to converge) or where the fits completed but the fit parameters were deemed unphysical and fell outside of the conditions set for good fits in Section~\ref{sec:success}, respectively.}
	\end{tabular}
\end{table}

In addition to the MaNGA datacubes, several Value Added Catalogs from the DR15 were used in this project, namely the MaNGA \textsc{PyMorph} DR15 Photometric Value Added Catalog \citep[MPP-VAC;][]{Fischer_2019} and the MaNGA Deep Learning Morphology Value Added Catalogue \citep[MDLM-VAC,][]{Dominguez_2018}. These catalogs will be described in more detail later in the paper.

\section{Modelling the galaxies with \textsc{buddi}}\label{sec:method}

Each galaxy datacube was modelled with \textsc{buddi} using both a single S\'ersic (SS) model and a S\'ersic+exponential (SE) model, where the latter may be interpreted as a fit to the bulge and disc respectively. \textsc{buddi} uses \textsc{galfitm} \citep{Haeussler_2013}, a modified form of \textsc{galfit} \citep{Peng_2002, Peng_2010}, to model the light profile of multi-waveband images of a galaxy simultaneously. The benefit of using \textsc{galfitm} over \textsc{galfit} is that the variations in the structural parameters of each component with wavelength can be modelled using user-defined Chebychev polynomials. As a result, the fit to each image slice uses information from the entire datacube, thus boosting the S/N over that of any individual image and allowing reasonable estimates to be derived for the structural parameters at wavelengths where the image slices have lower S/N. This technique is especially powerful in separating different components of different colours \citep{Haeussler_2022}, ideal to separate the spectra of bulge and disc. 
While we encourage readers to be familiar with \textsc{buddi}, where a full description can be found in \citet{Johnston_2017},  a brief overview and workflow are presented below, and visualized in Fig.~\ref{fig:buddi_sequence}.

\begin{figure*}
 \includegraphics[width=0.9\linewidth]{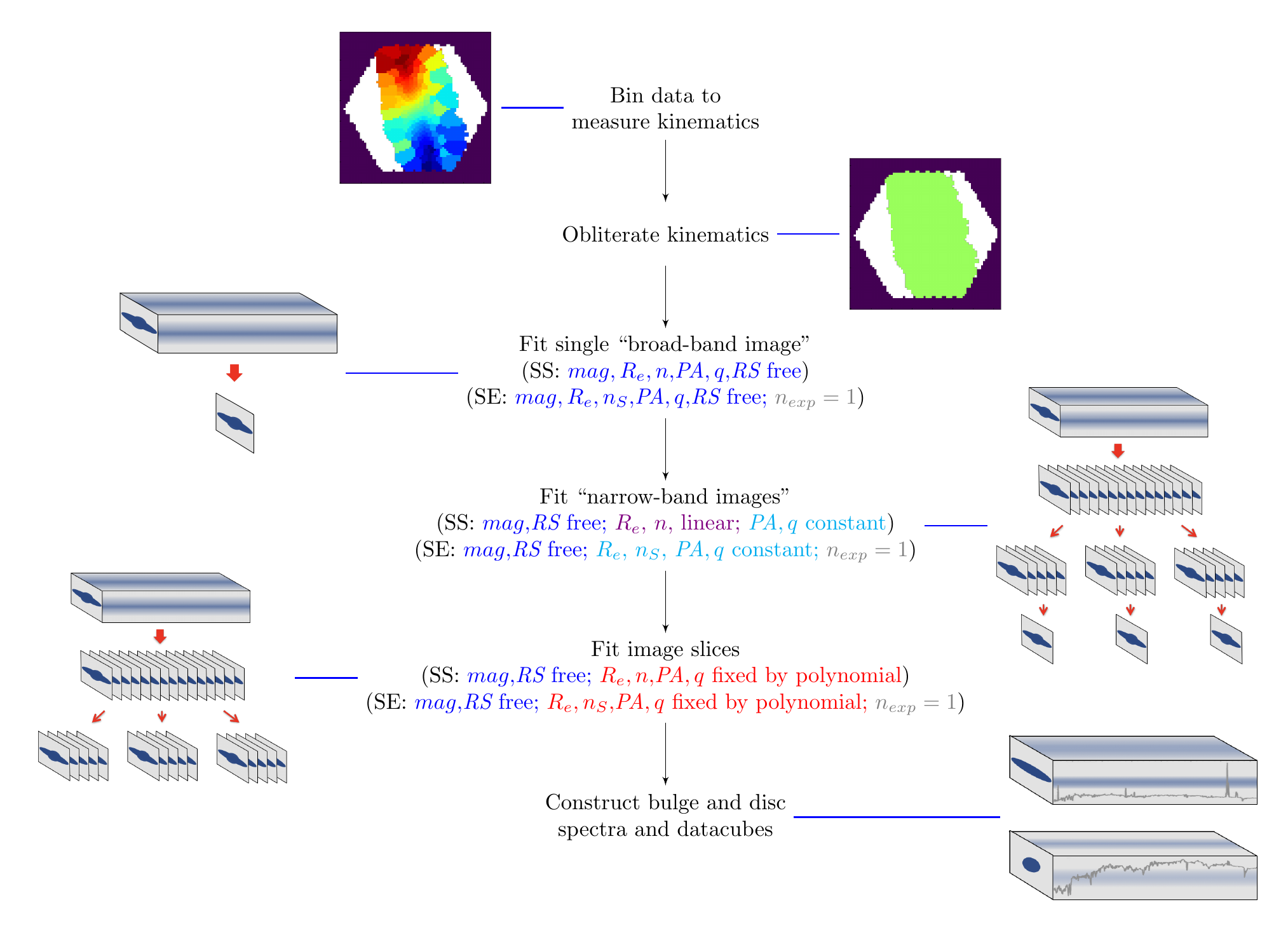}
 \caption{An overview of the steps to decompose a MaNGA datacube with BUDDI. The free parameters are the integrated magnitude ($mag$), effective radius ($Re$), S\'ersic index ($n$; in the SE fits, $n_S$ and $n_{exp}$ refer to the S\'ersic indices of the S\'ersic and exponential component respectively), position angle ($PA$), axes ratio ($q$) and residual sky value ($RS$). SS and SE refer to the single S\'ersic and S\'ersic+exponential fits, respectively, and the colours of each parameter identify how it varies with wavelength - dark blue means it’s completely free with wavelength, purple is allowed to vary according to a linear polynomial, light blue can be different for different components, but remains constant with wavelength, and red is held fixed according to the best fit polynomial from the previous step. }
 \label{fig:buddi_sequence}
\end{figure*}


\subsection{Preparation: Create the mask and PSF profile}\label{sec:step_0}
The first step towards carrying out the fits  with \textsc{buddi} is to prepare the data, including the corresponding Point Spread Function (PSF), bad pixel mask and  uncertainty, or sigma, datacubes. 

\subsubsection{Bad-pixel mask}
Since the MaNGA IFUs have hexagonal FOVs, the rectangular datacubes contain many pixels with 0 values filling in the space outside the MaNGA field. If left unmasked, \textsc{galfitm} will assume that these pixels are valid and represent the sky level, which will affect the fit due to the sharp cut off where the galaxies extend out of the MaNGA FOV. Consequently, the first step was to create a bad pixel datacube to mask out these spaxels, which was extracted from the mask extension within each MaNGA datacube. This extension gives details on the quality of each spaxel in the datacube, including information on the effects of the IFU footprint and missing data, and thus identifies which spaxels should be used or ignored as a function of wavelength in the analysis. Consequently the mask varies as a function of wavelength, and so the next steps carried out with \textsc{buddi} applied the same binning steps to the mask datacube that were used for the science datacube to ensure the correct spaxels were masked out in all steps of the fit. 

In the current version of the data, no further steps were taken to mask out other objects within the FOV, such as foreground stars or background/neighbouring galaxies, which may ultimately affect the final fits. However, this step is to be included in the next round of fits using the data released as part of DR17.

\subsubsection{PSF model}
Another part of the preparation was to create a model PSF datacube for each galaxy, which will be convolved with the fit to all images in the datacube to improve the fit. \citet{Yan_2016} describes how the IFU fibre bundles used by the MaNGA survey cannot alone provide an accurate PSF shape since some flux is lost in each exposure to the gaps between the fibres, resulting in it being spatially undersampled. While a three point dither pattern was used for all observations to provide full spatial coverage and to create a more uniform circular PSF \citep{Law_2015}, additional steps were taken to identify the shape of the PSF in order to refine the scale of the PSF observed with the fibre bundles.

To refine the shape of the PSF, the images of the guide stars were used. For each 15 minute exposure, a series of guider images were stored, and the integrated PSF was created by bias subtracting and flat fielding these images before stacking them. The resulting image is the time-integrated PSF, and includes any smearing introduced by imperfect guiding. The typical FWHM of this stacked PSF image is $\sim1.5\arcsec$. The MaNGA data reduction pipeline then modeled the focal-plane PSF observed by the fibre bundles with a double Gaussian profile, and used this profile to refine the shape of the PSF created from the guider images. The result is the reconstructed PSF, which is included in the datacubes for the $griz$-bands. According to \citet{Law_2016}, the MaNGA reconstructed PSF is nearly constant with wavelength, and the spatial covariance matrix also varies only slowly with wavelength. Therefore, the $griz$-band reconstructed PSF images describe the covariance in the data cubes by providing sparse correlation matrices in the $griz$-bands, from which the  PSF image at any wavelength can be created by interpolation. 

In this work, the PSF datacubes were thus created by interpolating between the reconstructed PSF images in the  $griz$-bands, using the filter transmission curves to determine the relative flux of each PSF image at each wavelength.

\subsubsection{Sigma datacube}
To further improve the fit, a sigma datacube was created from the inverse variance (IVAR) datacube included in the extension of the MaNGA datacube using $sigma=1/\sqrt{IVAR}$.  This datacube can be handed over to \textsc{GalfitM} as part of the fit, and allows more accurate estimation of the flux uncertainty in each pixel than an internal estimation by \textsc{GalfitM} from header keywords. \\

\noindent Finally, the target and sigma datacubes were converted from flux units of erg/s/cm$^2$/\AA\  into Janskys. This conversion results in the spectra having a constant magnitude zeropoint of 8.9 over the entire wavelength range, which is one of the input parameters for \textsc{GalfitM}, and allows greater flexibility when using \textsc{buddi} for data from different IFU spectrographs. Note that this conversion is reversed at the end of the process (See Section~\ref{sec:step_4}).


\subsection{Step 1: Obliterate the kinematics}\label{sec:step_1}
The first step that \textsc{buddi} carried out was to measure and obliterate the kinematics across the galaxy, thus ensuring that the image of the galaxy at each wavelength is at the same rest-frame wavelength and giving a symmetric light profile. This step is important because \textsc{buddi} uses  \textsc{GalfitM}  to model symmetric light profiles for each galaxy in this study. The kinematics of the galaxy would result in asymmetric images of the galaxy in the uncorrected datacube at wavelengths around strong spectral features due to the red and blue shifts of those features along the major axis of the galaxy. Additionally, the decrease in velocity dispersion with radius would contribute to further asymmetries in the light profiles of individual image slices, where the light in the inner parts of the galaxy may reflect the wings of a spectral feature while in the outskirts it comes from the stellar continuum. Fitting an uncorrected datacube of a galaxy with \textsc{buddi} results in spectral features similar to P~Cygni profiles in the final spectra of each component, and so these corrections are necessary for the quality of the final spectra.

In order to carry out these corrections, the datacubes were first binned using the Voronoi tessellation technique of \citet{Cappellari_2003}, and the kinematics of the binned spectra were measured with the penalised Pixel Fitting software (\textsc{ppxf}) of \citet{Cappellari_2004}. This latter step was carried out using the template spectra of \citet{Vazdekis_2010}, which are based on the MILES stellar library of \citet{Sanchez_2006}. They consist of 156 template spectra ranging in metallicity from $-1.71$ to $+0.22$, and in  age from 1 to 17.78 Gyrs. Having measured the kinematics across the galaxy, the spectra in each spaxel within the datacube were shifted to correct for the line-of-sight velocity, and broadened to match the maximum velocity dispersion measured within the galaxy.  To remove erroneous measurements due to the background noise, a limit of S/N~$=3$~per pixel at $\sim6000$\AA\ was implemented to only include those spaxels that were not dominated by noise.


\subsection{Step 2: Fit the white-light image}\label{sec:step_2}
Having obliterated the kinematics across the datacube, the light-profile fitting can start. The first step is to create the white-light image of the galaxy, and fit it using the SS and SE models, respectively. This step provides a quick way to determine the best initial parameters and number of components for the fits using an image with the maximum S/N. While this procedure is mainly useful for manual fitting of small samples of galaxies, it is still useful for the automated fits carried out in this work as it helps to refine the initial estimates for the fit parameters (integrated magnitude, $m_{tot}$; effective radius, $R_e$; S\'ersic index, $n$; axis ratio, $q$; and position angle, $PA$) before introducing the wavelength information. The initial parameters for these two fits were taken from the MPP-VAC, in which the $g$, $r$ and $i$-band images of the galaxies included in the DR15 MaNGA sample from the NASA-Sloan Atlas catalog \citep[hereafter NSA,][]{Blanton_2011} were modelled with the \textsc{PyMorph} code \citep{Vikram_2010}, which also uses \textsc{Galfit}, albeit single-band. The MPP-VAC carried out these fits for both SS and SE models. Therefore, for each galaxy, the $r$-band parameters from the MPP-VAC for the corresponding model (SS or SE) were used as initial parameters for the fits with \textsc{buddi} since this band falls closest to the centre of the MaNGA spectra wavelength range. Additionally, the effective radius measurements from the MPP-VAC were corrected for the pixel scale of the MaNGA datacubes, and the position angle from the MPP-VAC was corrected to agree with how \textsc{Galfit} measures the angle.  

The MPP-VAC includes flags to identify failed fits within the sample for either the SS or SE models, which usually occurred for galaxies with irregular morphologies, in the process of merging or where multiple galaxies are superimposed within the FOV. Consequently, only those galaxies marked as having a successful fit in the MPP-VAC were modelled with \textsc{buddi}.  This additional constraint reduced the total number of candidate galaxies that could be fitted to 1,710 for the SS model, and 1,705 for the SE model (listed as Number of candidate galaxies (MPP-VAC) in Table~\ref{tab:numbers}). The fits with each of these two models were carried out independently, resulting in some galaxies that were modelled successfully with one model but not the other.


\subsection{Step 3: Fit the narrow-band images}\label{sec:step_3}
Having carried out the fit to the white light image, the next step is to introduce variations due to wavelength. If the fit to the white-light image was successful (i.e. \textsc{GalfitM} converged on a solution and didn't crash), the datacube was rebinned into a series of 10 narrow-band images along the wavelength direction, where these images have higher S/N than any individual image slice within the datacube. While the user can define the number of narrow band images, we decided to use 10 since it gave a good compromise between the speed of the fits (more images means more free parameters and slower fits) and a good coverage of the wavelength range. The initial estimates for the fit parameters for these images were taken as the fit parameters from the fits to the white light images in the previous step.  In each case (SS and SE), the first fit to these narrow band images was allowed complete freedom in the fit parameters, simulating the results obtained by running \textsc{Galfit} on each image independently. This step gives a good idea of how the structural parameters vary as a function of wavelength. In manual fits for small samples of galaxies, these results would be useful for determining the correct Chebychev polynomials for each parameter to reflect the general variation as a function of wavelength. In these automated fits however, the Chebychev polynomials are predefined. But it was still decided to run these free fits since they may  be useful for identifying galaxies that are not so well modelled with these set polynomials for a future follow-up study.

Having completed the free fits, the fits were then repeated using Chebychev polynomials to constrain the parameters for effective radius ($R_e$), S\'ersic index ($n$), axis ratio ($q$) and position angle ($PA$) as a function of wavelength while leaving the integrated magnitude ($m_{tot}$) free.  In the SS fits, a polynomial of order 2 was selected for the $R_e$ and $n$  and an order of 1 used for the $PA$ and $q$, following the approach of \citet{Vulcani_2014}, while in the SE fits, polynomials of order 1 (constant with wavelength) were used for these parameters. In both cases, the integrated magnitude was allowed full freedom. This additional degree of freedom in the sizes and S\'ersic indices for the SS fits effectively mimics colour gradients across the galaxy and allows for a better single component fit in the case of 2-component galaxies which would be more bulge dominated in the inner regions and more disc dominated in the outskirts. In such cases, the galaxy would appear larger and with a lower S\'ersic index at bluer wavelengths, and more compact and potentially with a higher S\'ersic index at redder wavelengths.  Thus, allowing these variations with wavelength enables GalfitM to derive a more accurate fit to the light profile of the galaxy as a whole.

\subsection{Step 4: Fit the image slices and extract the decomposed spectra}\label{sec:step_4}
Once the structural parameters have been determined in the narrow band images, it is now possible to determine the magnitude/flux at each wavelength, which ultimately will be used to derive the spectra for each component. This step was carried out by fitting the unbinned image slices from the kinematics-obliterated datacube. In order to reduce  the computing time  and to avoid memory issues, each datacube was split up into batches of 10 consecutive image slices, which were modelled by \textsc{GalfitM} simultaneously. In these fits, the structural parameters ($R_e$, $n$, $PA$, and $q$) were held fixed for each image slice, using the values that were determined in the previous step, while the integrated magnitude was left completely free.  The initial estimates for the magnitudes for  each image slice were determined by interpolation at the corresponding wavelength between the fits to the binned images. It should be noted at this point that no special treatment was applied for galaxies containing emission lines. Since the parameters were held fixed according to the polynomials used, any emission lines that displayed a similar light profile to one of the components was included in the fit, and thus the spectrum, for that component. If however the emission lines followed a different distribution, \textsc{GalfitM} would still model the underlying galaxy and the emission line features would appear in the residual images and datacube. Tests were carried out comparing the stellar populations derived from the S\'ersic and exponential component spectra with those from the original datacube in galaxies that contain emission lines, and no significant difference in the final results was seen. However, this topic will be explored in greater detail when we repeat the fits to the DR17 dataset.


\begin{figure*}
 \includegraphics[width=0.9\linewidth]{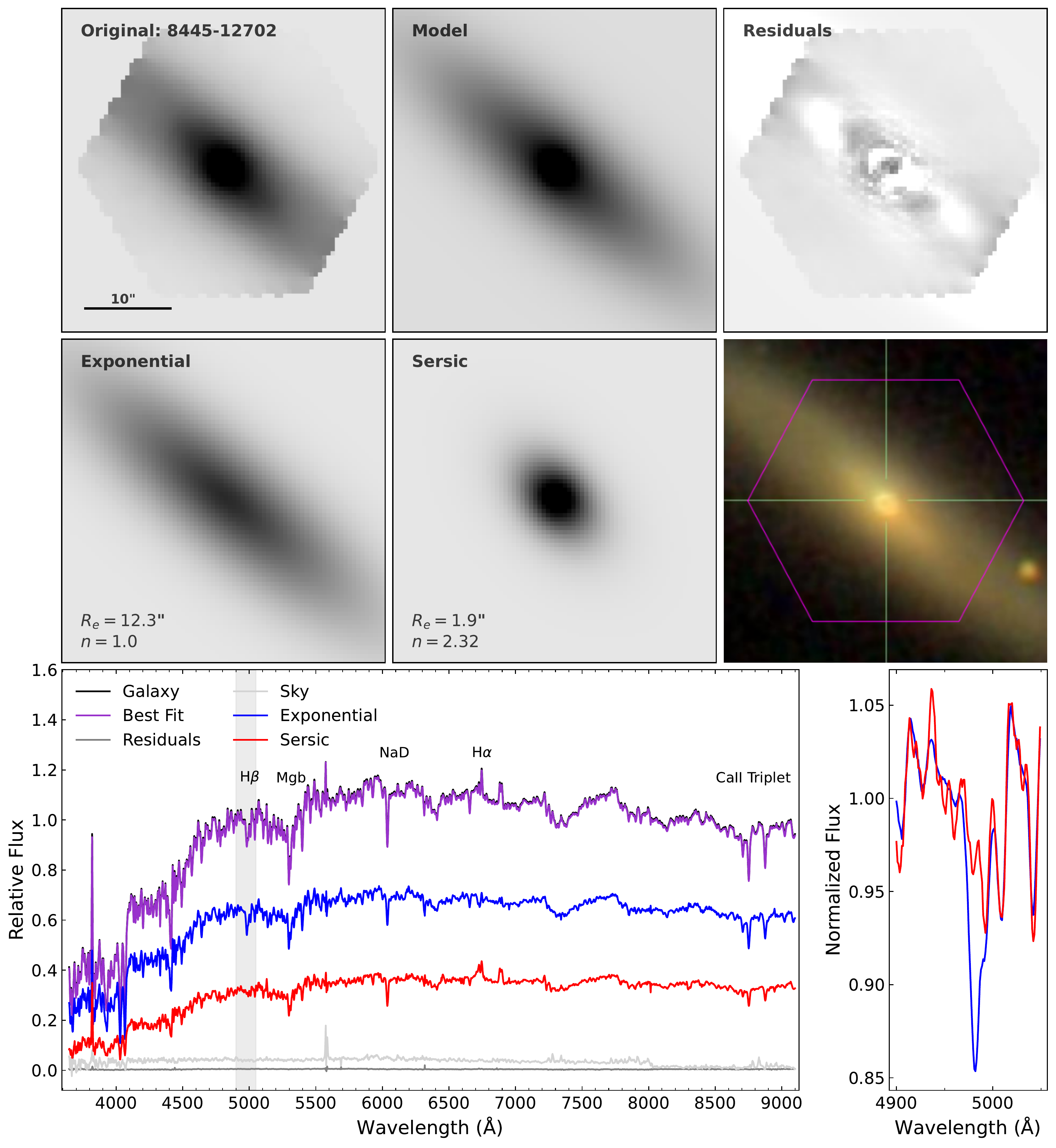}
 \caption{Top row: the MaNGA white-light image of 8445-12702, the best fit model for the SE fit, and the residual image. Middle row: the model images for the exponential and S\'ersic components, including information on the $R_e$ and $n$ from the fits, and the SDSS $gri$-band image, downloaded using the Marvin Python client \citep{Cherinka_2019}. All images from the MaNGA data have been scaled to the same flux range for easier comparison. Bottom row: The normalized integrated flux of the galaxy (black) from the MaNGA datacube, with the spectra extracted for the exponential  and S\'ersic components in blue and red, respectively. The dark and light grey lines represent the residuals from faint asymmetric features and background objects and the sky background, respectively. The feature in the sky spectrum at $5577$\AA\ is the [OI] sky line. The purple line represents the combined exponential+S\'ersic+sky spectrum, superimposed upon the integrated spectrum of the galaxy. On the right is a zoom-in of the normalised exponential and S\'ersic spectra over the spectral region marked by the grey band on the left plot, showing the differences in the line strengths of the H$\beta$ feature between the discs and the bulge.  }
 \label{fig:spec_1}
\end{figure*}

Once the fits to these image slices were completed, the spectra for each component were created in 2 ways. First, as one-dimensional integrated spectra  for the galaxy component as a whole, produced by plotting the integrated flux (derived from the magnitude from \textsc{GalfitM}) of that component as a function of wavelength, and second as model datacubes, created using the best-fit model images of each component at each wavelength from \textsc{GalfitM}. In both, the fluxes of these spectra were converted from Janskys back into the original flux units.  An example of the fit to galaxy 8445-12702 and the spectra extracted for the S\'ersic and exponential components are shown in Fig.~\ref{fig:spec_1}, along with a zoom in on the H$\beta$ feature to show the differences in the final spectra. The images created from the MaNGA data have all been scaled to the same flux levels for easier comparison, and the residual image reveals faint spiral arms within the disc of this galaxy that cannot be seen in the original MaNGA datacube or the SDSS $gri$-band image.

The results presented and discussed in the rest of this paper use the one-dimensional integrated spectra. Due to the polynomials used in the fits to the image slices, the decomposed datacubes show the same spectrum, and thus stellar populations, in every spaxel. Therefore, while these datacubes do not show any variations across the bulge or disc components, they can still be useful for comparing the bulge and disc luminosity fractions across the galaxy and thus better understanding the stellar population gradients seen within a galaxy (e.g. see Figures~14, 15 and 16 of \citealp{Johnston_2017}, and Figure~7 of \citealp{Johnston_2018}). Additionally, they can be used to create a residual datacube that is free from the light of the galaxy, which can be useful for studying faint background or neighbouring sources whose light would otherwise be contaminated by the light of the main galaxy (e.g. see Solimano et al, submitted).


\section{Overview of the fits}\label{sec:overview}
As described in Section~\ref{sec:data}, 1926 candidate galaxies were identified from the total sample included in the DR15 by selecting only those observed with the 91 and 127-fibre IFUs. However, not all of these galaxies had successful fits, and consequently fit parameters, in the MPP-VAC. By including only those galaxies with successful fits in the MPP-VAC, the sample was reduced to 1719 and 1705 for the SS and SE fits, respectively.  Of these galaxies, good fits with \buddi\ were achieved for 1038 (SS)  and 691 (SE) galaxies, corresponding to a success rate of $\sim70\%$ and $53\%$, respectively. The remaining galaxies fell into two categories- those in which the fit with \textsc{GalfitM} crashed and/or failed to converge (see Section~\ref{sec:crashed}), and those that were considered poor fits due to unphysical structural parameters (see Section~\ref{sec:success}). An overview of these numbers is given in Table~\ref{tab:numbers}.
 
In this section, we will explore the properties of the fits, the criteria used to identify a good fit, and the reasons for some fits failing.


\subsection{Common reasons for failed/crashed fits}\label{sec:crashed}
In around $23\%$ of the model fits by \textsc{GalfitM} to the candidate galaxies (i.e. those with successful fits in the MPP-VAC), the fits crashed or failed to converge, either when modelling the white-light image, the narrow band images, or the image slices (see Sections~\ref{sec:step_2}, \ref{sec:step_3} and \ref{sec:step_4}, respectively).  There are several reasons for these fits to fail, which are discussed below.

One reason for many of the failed fits is due to the small FOV of the MaNGA images compared to the SDSS imaging data used for the fits in the MPP-VAC. Since the MaNGA galaxies are only observed out to 1.5 or 2.5~$R_e$, limited information is available for the outskirts of the galaxies. This issue becomes particularly problematic for more face-on galaxies, where the MaNGA FOV has few or no pixels representing the light from the outskirts of the galaxy or purely the sky background. As a result, in these cases, the fit fails due to insufficient information being available.

Another issue is that foreground and background objects that fall within the MaNGA FOV are not masked out in the fits with \textsc{buddi}- the masks created only represent the spaxels within the datacube that contain no light from the IFU fibres. Any bright, unmasked objects in the field can affect the fits by distorting the shape of the model created by \textsc{GalfitM}. While this issue is limited by the small FOV of the MaNGA datacubes, the effect is not negligible. In a future version of the fits to the full MaNGA sample, the masking algorithm will be improved to identify such contaminating sources and to mask them out.

While the MPP-VAC gives a flag for successful fits, which were used as initial estimates for the fits by \textsc{buddi}, in some cases the fit parameters given in the MPP-VAC were unphysical. For example, a small number of galaxies showed axes ratios for one component as being $<0.1$, or having very small effective radii (e.g. less than half a pixel width in the SDSS imaging data). Consequently, in some cases, using these unphysical parameters as initial estimates for the fits with \textsc{buddi} resulted in some of the fits failing.

Finally, another reason for some fits failing is due to the models either containing too many/few components. For example, attempting to fit a galaxy with a complex morphology, such as with a bulge, disc, bar and spiral arms, with the SS or SE models may result in a poor fit due to the complexity of the galaxy. A similar but related issue is caused where one component is particularly faint at specific wavelengths. For example, in some galaxies it was found that the bulge was very bright at redder wavelengths but very faint in the blue end of the spectrum. As a result, the fits to the image slices failed at bluer wavelengths since the bulge was too faint to be reliably detected, but were successful at redder wavelengths where the bulge was brighter.

While the reasons outlined above account for the common issues that led to the majority of the crashed fits, they do not represent an exhaustive list of all the possible reasons that caused the fits to fail.


\subsection{Classification of a good fit}\label{sec:success}
Of the fits that completed successfully and provided spectra for each component included in the models, a further selection was applied to distinguish between good (i.e. reliable) and bad (unreliable)  fits. This selection criteria was based on the feasibility of the final structural parameters of the fit, where those that converged on unphysical parameters were considered bad fits, and were applied to the fits to the narrow-band images. The following criteria were used to identify good fits:
\begin{enumerate}
 \item $\Delta mag_r \leq 2.5$ (SE models only) 
 \item $0.25 \leq R_e \leq 50\arcsec$,
 \item $0.205 \leq n \leq 7.95$,
 \item $0.1 \leq q \leq 1.0$.
\end{enumerate}

In the SE models, if only one component was found to fall outside of these parameters, the fit to the whole galaxy was considered bad. The limit on the magnitude difference in the SE fits was selected since it represents a case where the minimum light fraction of both components is $10\%$. A visual inspection of the fits showed that in many cases where $\Delta mag_r > 2.5$, the galaxy was well modelled with a single S\'ersic profile. Thus, in these cases, the SE fits `failed' because the galaxies are single component systems, i.e. they do not contain a distinct bulge and disc.

The lower limit for the $R_e$ corresponds to half a spaxel in the MaNGA datacube ($\sim10\%$ of the PSF FWHM), which was considered the minimum reasonable size that could be determined with \textsc{galfitm}. \citet{Haeussler_2013} and \citet{Haeussler_2022} have shown that, if the PSF is well known, recovering a object size well below 1 pixel is feasible without systematic bias (down to 0.3~pixels for one-component fits in \citealt{Haeussler_2013} and 0.5~pixels for two-component fits in \citealt{Haeussler_2022}). From fits to single band images, \citet{Gadotti_2008} recommended a lower limit of $R_e$ of 0.8 times the half width at half maximum (HWHM) of the seeing PSF (i.e. 0.6\arcsec\ in this work). We therefore looked at the effect on the final sample using different lower limits to the $R_e$, finding no cases with $R_e < 0.6\arcsec$ for the SS models, and only 1 exponential and 8 S\'ersic components with $R_e < 0.6\arcsec$ for the SE fits (with the majority of these still having radii of greater than 0.5\arcsec). Since so few galaxies are affected by increasing the lower limit, we  decided to use the lower limit for $R_e$ based on the multi-band fits of \citet{Haeussler_2013,Haeussler_2022}.
The upper limit in $R_e$ is larger than the FOV of all MaNGA IFUs, and was selected to try to maximise the sample of good fits while eliminating any obviously unreliable fits.  In \citet{Johnston_2017}, It was found that for more edge on galaxies observed with MaNGA, \textsc{galfitm} can overestimate the effective radius, sometimes to a value outside of the radius of the FOV, since only the inner regions and the minor axis of the galaxy falls within the FOV. However, despite this issue, the residual images show a very good fit to the galaxy within the FOV, and so the spectra extracted for both components were  considered reliable. However, in these cases in particular, it should be noted that the resulting spectra will only represent the inner regions of the galaxy within the FOV available. 

The fits were carried out using a range of S\'ersic indices of between 0.2 and 8, which are the typical upper and lower limits generally imposed upon similar fits in the literature. In order to avoid using any models that hit these hard limits and thus failed to converge, we followed  \citet{Haeussler_2022} to use the range $0.205 \leq n \leq 7.95$. This limitation had the largest effect on the number of galaxies considered to have a `good' fit, and in the next version of the fits to the DR17 data we plan to revisit the fits that hit these limits to try to improve the models using different starting parameters or a narrower range for the S\'ersic index.

The lower limit on the axes ratio rules out poor fits due to unmasked foreground stars within the FOV that \textsc{galfitm} attempted to include in the fit. This limit are in alignment with the sample selection carried out in \citet{Haeussler_2013} and other works using \textsc{GalfitM}.

$\sim7\%$ and $\sim23\%$ of the fits to candidate galaxies resulted in bad fits based on the criteria above for the SS and SE models, respectively.  It was found that in general, the fits that resulted in unphysical parameters were either nearly face-on galaxies, where it was difficult for \textsc{galfitm} to reliably model the sky background or the outskirts of the galaxy, or those that showed other objects  within the MaNGA FOV, such as foreground stars, neighbouring/background galaxies, bright spiral arms or star-forming regions. While some of these cases resulted in the fits crashing (see above), others led to distorted models for the galaxy and unphysical structural parameters.

The presence of bars should also be taken into account when considering the fits as this component may affect the fit to the bulge, thus affecting the structural parameters and the derived spectra. While a full analysis of the effect of bars on the fits is beyond the scope of this paper, we can however look at the expected fraction of bars in the sample of galaxies. The MDLM-VAC gives the probability for each galaxy containing a signature of a bar from Galaxy Zoo 2 \citep{Willett_2013} and \citet{Nair_2010}, which uses data from the SDSS DR4 \citep{Stoughton_2002}. By adopting a limit of 0.5 for the presence of a bar, we find that 112 and 126 galaxies successfully fitted with the SE model have a clear bar signature according to Galaxy Zoo 2 and \citet{Nair_2010}, respectively. Of these galaxies, 72 galaxies are in common between the two catalogs, accounting for $\sim10$~percent of the SE sample of galaxies with successful fits. Thus, the majority of the galaxies modelled with the SE model in this work can be considered unbarred.

A summary of the number of candidate galaxies versus the number of galaxies that were modelled successfully with \textsc{buddi} is given in Table~\ref{tab:numbers}. Throughout the rest of this paper, the discussion will focus on the successful fits using the criteria above, unless specified otherwise. It should be noted at this point that the classification of successful fits does not take into account the S/N of the extracted spectra, and future works using the BUDDI-MaNGA data may choose to use their own selection criteria depending on their science goals.


\begin{figure*}
 \includegraphics[width=1\linewidth]{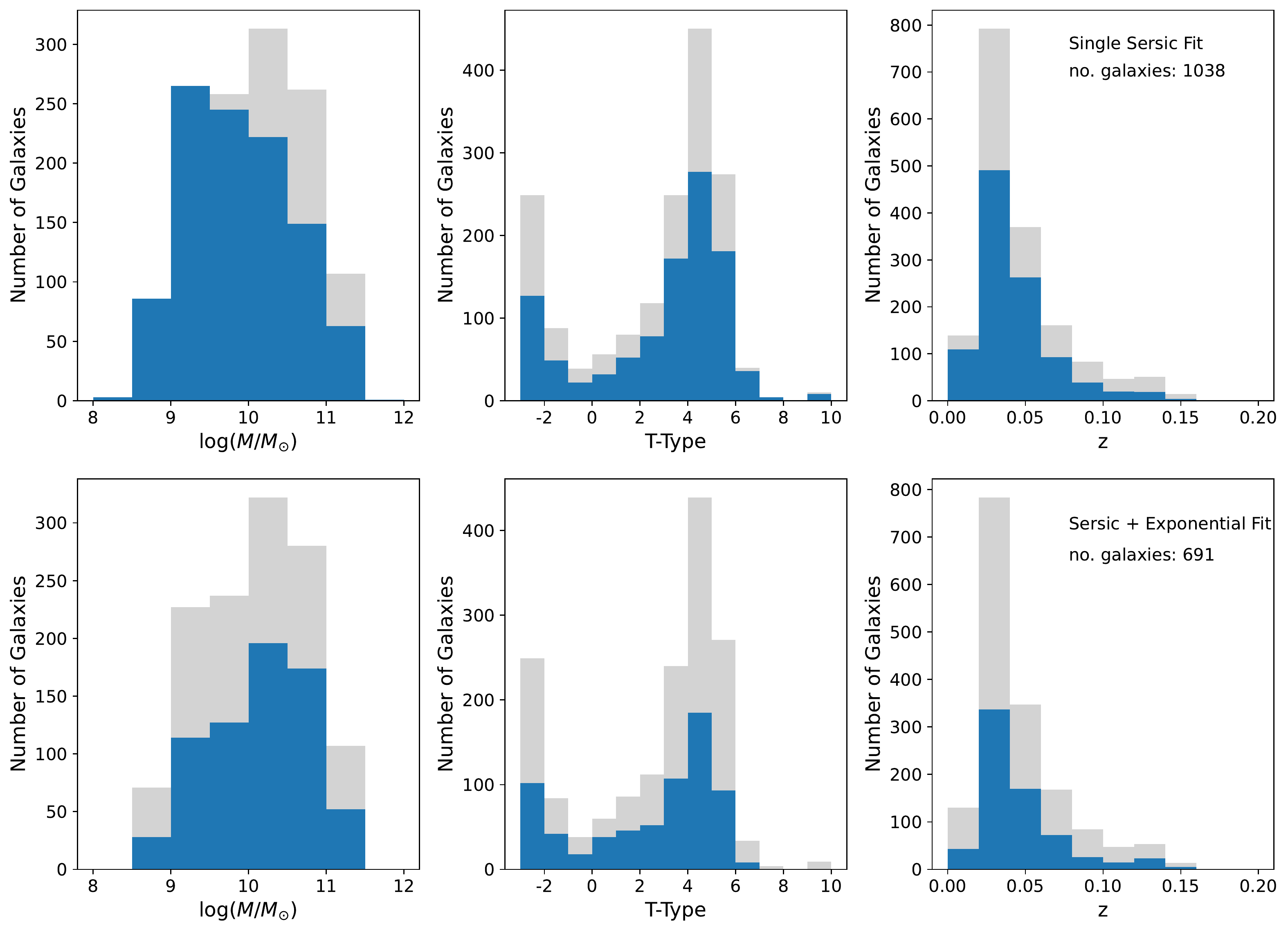}
 \caption{Overview of the masses (left), morphologies (middle) and redshifts (right) for `good' fits with the SS (top) and SE (bottom) models. The grey bars represent the morphologies of all galaxies where fits where attempted (i.e. observed with the 91 and 127-fibre IFUs with successful fits in the MPP-VAC for SS or SE profiles). 
 }
 \label{fig:distribution_mass_morph_z}
\end{figure*}

\subsection{Distributions of the successful fits}\label{sec:sample_classification}

Having identified the galaxies with successful fits within the sample, we can now explore the properties of those galaxies components. The distribution of the galaxies with successful fits in terms of total mass, morphology and redshift are shown in Fig~\ref{fig:distribution_mass_morph_z}. The masses and redshifts were taken  from the NSA, while the morphologies come from the MDLM-VAC, where early type galaxies (ellipticals and S0s) are considered to have T-Type~$<0$ while late type (spiral) galaxies have T-Types~$>0$ and irregular galaxies are identified with T-Type~$=10$. The figure also shows the distribution of the full sample of MaNGA candidate galaxies from DR15 (i.e. those with initial estimates for the fit parameters from the MPP-VAC and observed with the 91 or 127-fibre IFUs) in grey for comparison.  It can be seen that the distribution of the galaxies successfully modelled by \textsc{buddi} do follow approximately the same shape as the full DR15 sample, with approximately $40-50$\% of galaxies in each bin successfully modelled in SE, with significantly more in SS. The morphologies show peaks at T-Types -2.5 and 4.5, corresponding to early and late types respectively, and do not appear to show any bias against a certain T-Type, and the redshifts rise sharply to $z\sim0.03$ with a longer tail out to $z\sim0.15$. The shapes for the distributions for the SS and SE fits are very similar for morphology and redshift, but they show a larger difference in the galaxy stellar mass, where the SS models successfully fit more lower mass galaxies ($\text{log}(M/M_\odot)<10$) than the SE fits. One contributing factor to this effect could be that the lower mass galaxies were observed with the smaller IFUs, in this case the 91-fibre IFU, where it's harder to achieve a good fit with 2 components. Another consideration is that, in general, it is harder to separate the components in fainter galaxies due to lower S/N.

However, in general, the sample of DR15 galaxies that have been successfully modelled with \textsc{buddi} follow similar distributions in terms of the masses, morphologies and redshifts as the full MaNGA DR15 sample, indicating that these fits have introduced no significant bias to the sample.


\subsection{Statistics of the fit parameters}\label{sec:fit_params}

\begin{figure*}
 \includegraphics[width=0.9\linewidth]{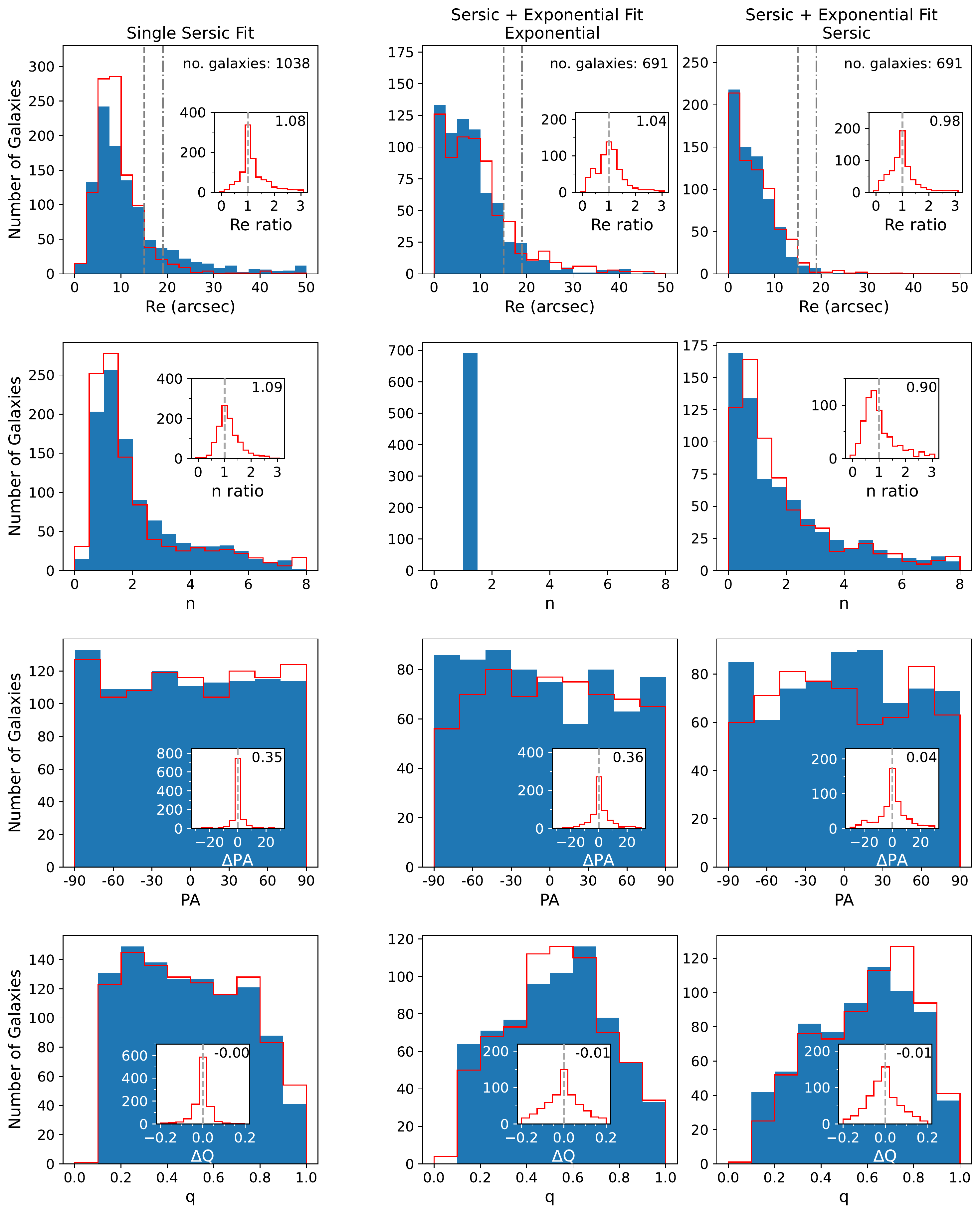}
 \caption{Overview of the fit parameters for `good' fits, defined as having $0.25<R_e<50\arcsec$, $0.205<n<7.95$ and $q>0.1$. The red line represents the distribution in the fit parameters from the MPP-VAC for the same sample of galaxies for comparison. The vertical dashed and dot-dashed lines in the top row indicate the radii of the FOV for the 91 and 127-fibre IFUs respectively. The insets in each plot show the histograms for the ratio of the $R_e$ and $n$ from \textsc{buddi} relative to the MPP-VAC values for the same galaxy and component (top two rows), and the differences in the \textsc{buddi} and MPP-VAC values for the $PA$ and $q$ in the bottom two rows. The numbers in the insets reflect the median value for the distributions in those histograms (i.e. the median value for the ratios or differences), and the dashed grey lines reflect ratios of 1 and differences of 0 as a guide.}
 \label{fig:distribution_fit_params}
\end{figure*}

The histograms in Fig.~\ref{fig:distribution_fit_params} present the distribution for each of the structural parameters ($R_e$, $n$, $PA$ and $q$) for all the successful fits with \textsc{buddi} in blue, with the distribution of the initial estimates from the MPP-VAC for the same galaxies in red for comparison. The distributions for $R_e$ in all plots show a peak at $R_e<10\arcsec$ with a long tail to larger radii. This trend is likely to be due to a selection bias in the original sample, where the IFU-bundle used to observe each galaxy was selected to cover the inner 1.5-2.5~$R_e$. As a reference, the vertical dashed and dot-dashed lines in these plots mark the radii of the 91 and 127-fibre IFUs, respectively. The fits to some of the galaxies with large values of $R_e$ were checked by eye, and it was found that many of them have high inclinations and appear very edge-on, making it hard to accurately measure the $R_e$ due to the small FOV. In the SE fits, it can be seen that the mean $R_e$ value for the exponential components is higher than the S\'ersic  component. Assuming these components model the discs and bulges respectively, this result reflects the scenario where the discs are more extended and the bulges more compact.

The S\'ersic index distributions are also interesting, showing a bias  towards lower values in the SS plot that peak at around $n=1$. The distribution in the morphologies in Fig.~\ref{fig:distribution_mass_morph_z}  shows that the majority of the sample are late-type galaxies (T-Type~$>0$), thus explaining the bias towards lower $n$. However, the fits to the S\'ersic component in the SE fits also show a peak at very low $n$. There are several possible reasons for this trend. For example, it could be due to the majority of bulges in this sample of galaxies being closer to pseudobulges than classical bulges \citep[using $n<2.0$ and $n>2.0$ respectively, using the definition of][]{Fisher_2008}, or that \textsc{GalfitM} has `flipped' the components, such that the exponential profile fits dominates the light in the inner regions and the S\'ersic profile models the more extended component. This phenomenon will be explored further in Section~\ref{sec:flipped}.

The distribution in the $PA$ is relatively flat for both sets of fits, as expected, and the axis ratio ($q$) for the SS fits are relatively flat up to $q\sim0.8$, where they start to drop. While this result may reflect the sample selection in the larger DR15 sample of galaxies, another significant contributing factor is that it's harder to get a good fit to more face-on galaxies in the MaNGA sample due to the small fields of view, resulting in the outskirts of the galaxy and the sky background not being included in the fit. Another factor to consider is that determining axis ratios for more face on galaxies is harder than more edge on galaxies, i.e. the difference from axes ratios of 0.9 and 1.0 is much smaller than from 0.05 to 0.15. The SE fits, on the other hand, show a peak at $q\sim0.65$ for both components, with the fits to the exponential component dropping faster to higher values of $q$ while the S\'ersic profiles drop more steeply to lower values for $q$. Again, assuming that these two profiles generally model the discs and bulges respectively, these trends simply reflect that the discs have higher ellipticities while the bulges appear rounder, as would be expected. 

\subsection{Comparing the fit parameters with the MPP-VAC}\label{sec:comparison_MPPVAC}
Since the MaNGA images have a smaller FOV and slightly lower spatial resolution than the imaging data used for MPP-VAC, the fit parameters for each galaxy derived by both methods may be different because \buddi\ has less information to model the outskirts of the galaxy and the sky background. To explore the effect of this smaller field of view on the reliability of the fits, \citet{Johnston_2017}  created a series of simulated datacubes based on the MaNGA data for different IFU sizes, and ran these datacubes through \textsc{buddi} using an SE model to compare the derived structural parameters and stellar populations with those used to create the datacubes. They found that for the datacubes based on the 127 and 91-fibre IFUs, the  final structural parameters and the stellar populations derived from the resulting spectra were consistent with the properties that went into the model datacubes. They therefore concluded that these two IFU sizes in the MaNGA data have sufficient resolution elements to obtain a reliable two component fit. In the smaller IFU sizes however, the structural parameters and, in particular the stellar populations, were found to become increasingly unreliable. However, those tests were run on a small number of simulated datacubes. With the data set used in this work, it is possible to repeat these tests in a more statistically significant way by comparing the fit parameters to those derived from photometric data and provided in the MPP-VAC.

The red lines in the histograms in Fig.~\ref{fig:distribution_fit_params} represent the distributions in the fit parameters from the MPP-VAC for the same sample of galaxies. In general, the distributions are very similar, indicating that in the majority of these cases the MaNGA FOV is sufficiently large to achieve a good fit with \textsc{galfitm}.

\begin{figure*}
 \includegraphics[width=0.9\linewidth]{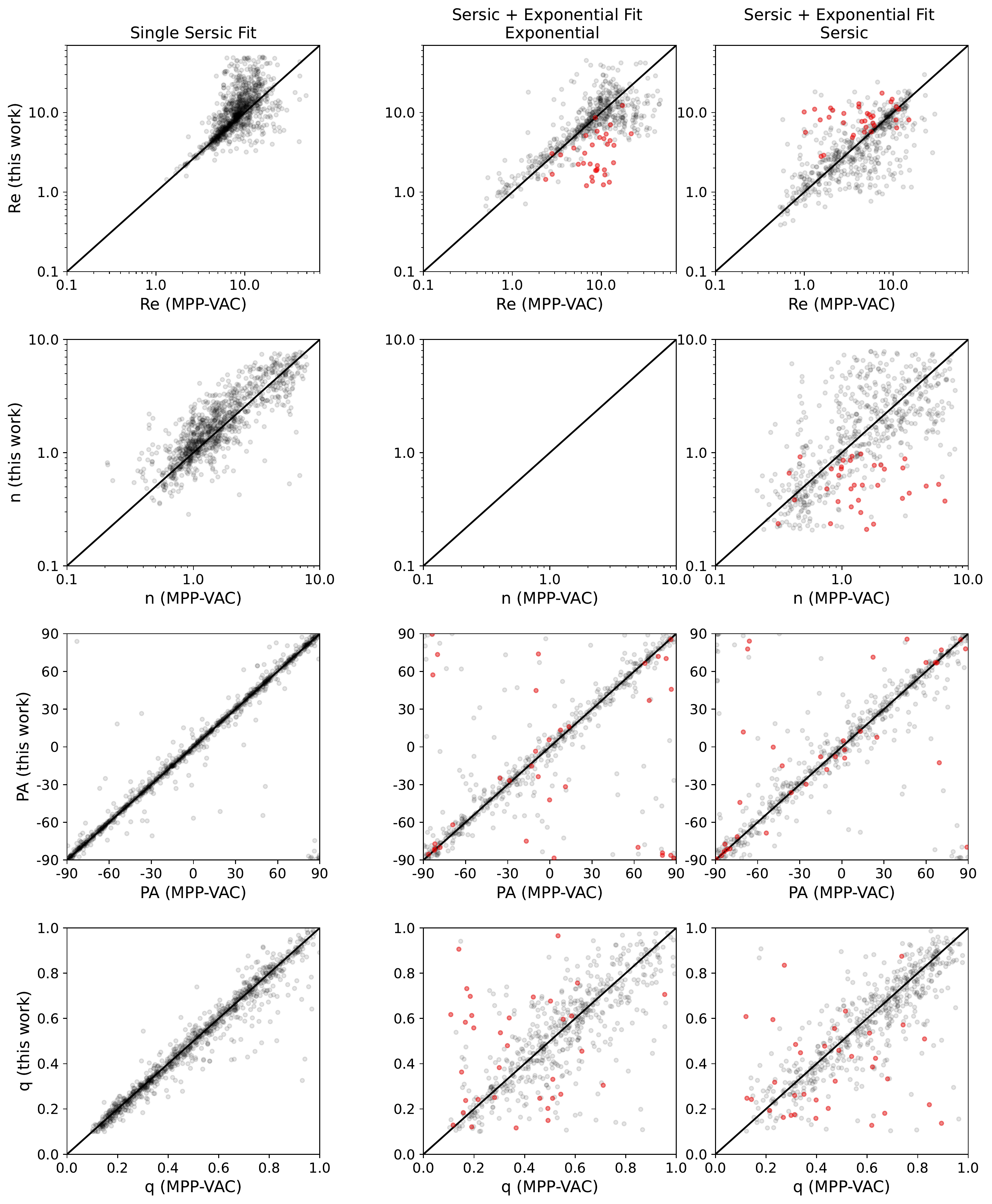}
 \caption{A comparison of the fit parameters derived using \textsc{buddi} with the input parameters from the MPP-VAC for galaxies considered to have `good' fits in both catalogs. Red points reflect galaxies identified as having been flipped by \textsc{buddi} but not in the MPP-VAC. The layout follows Fig.~\ref{fig:distribution_fit_params}}
 \label{fig:catalog_comparisons}
\end{figure*}

To further explore the differences in the fit parameters for each component modelled within each galaxy, the insets in each plot within Fig.~\ref{fig:distribution_fit_params} directly compare the parameters between \textsc{buddi} and the MPP-VAC for each galaxy, plotting the ratio of the $R_e$ and $n$ from \textsc{buddi} over the input value from the MPP-VAC in the top two rows, and the physical difference between the two fits for the $PA$ and $q$ in the bottom two rows. The median values for each distribution are given in the top right of each inset, and are listed in Table~\ref{tab:distributions} along with the values of the first and third quartiles of the distributions. It can be seen that in all cases, the distributions peak at around 1 for the ratios, and around 0 for the differences, indicating that, in general, the fits by \textsc{buddi} are not too different to the initial parameters taken for that galaxy or component from the MPP-VAC. Figure~\ref{fig:catalog_comparisons} also presents the fit parameters from \textsc{buddi} and the MPP-VAC as scatter plots. Again, it can be seen that the fit parameters are generally consistent, but with larger scatter for the $R_e$ and $n$. For example, one can see that \textsc{buddi} appears to over-estimate the $R_e$ for the single S\'ersic fits to larger galaxies, which is likely to be an effect of these galaxies filling more of the FOV and being poorly modelled by a single component. The S\'ersic indices for the S\'ersic components in the SE fits also display large scatter. Again, this trend may be an artefact from the effects already mentioned, but in some cases they may also be the result of the components being flipped by \textsc{buddi} and not in the MPP-VAC, or vice versa. In other words, the S\'ersic component may reflect the more extended component in the \textsc{buddi} fit but the more compact component in the MPP-VAC. This phenomenon will be discussed further in Section~\ref{sec:flipped}, but for now those galaxies that have been identified as flipped by \textsc{buddi} but not in the MPP-VAC are highlighted in red in Fig.~\ref{fig:catalog_comparisons}, and do appear to account for some of the scatter at low $n$ with \textsc{buddi}. Another effect may be the different approaches used in this work and the MPP-VAC for fitting galaxies with high S\'ersic indices, where  in the MPP-VAC, the authors repeated the fits to the bulge components that converged on $n=8$ with a reduced upper limit on the S\'ersic index constraining the fit, generally with $1 < n_{lim} < 3$. They then evaluated each fit in terms of the $\chi^2$ of the final fits, and selected the better fit, which was often the one with the lower limit for $n$. This process of repeating the fits explains the higher frequencies for $n\sim1$ in the MPP-VAC catalog seen in Fig.~\ref{fig:distribution_fit_params}, but may introduce potential bias in the fits to artificially low S\'ersic indices and leads to an inhomogeneous fitting sample due to the use of different constraints. Finally, while \citet{Haeussler_2013} and \citet{Haeussler_2022} show that multi-waveband fits, such as those used in this study, allow sizes and S\'ersic indices to be recovered more accurately, they still find that the S\'ersic index is the hardest parameter to derive accurately when applying light profile fits for both one and two-component models. Therefore, the scatter seen in the S\'ersic indices in Fig.~\ref{fig:catalog_comparisons} is to be expected.


\begin{table}
	\centering
	\caption{Overview of the distributions in the ratios of the $R_e$ and $n$ from \textsc{buddi} relative to the MPP-VAC values for the same galaxy and component, and the differences in the \textsc{buddi} and MPP-VAC values for the $PA$ and $q$, giving the median values and the values of the first and third quartiles.}
	\label{tab:distributions}
	\begin{tabular}{lrrr} 
 \hline
  			& Q1		& Median	& Q3 \\
 \hline
  \textbf{SS}	&		&		& \\
  $R_e$		& 0.96	& 1.08 	& 1.48\\
  $n$			& 0.90	& 1.09 	& 1.37\\
  $PA$ 		& -0.45 	& 0.35 	& 1.05\\
  $q$  		& -0.02 	& 0.00 	& 0.01\\
 \hline
  \textbf{SE, exponential}		&		&		& \\
  $R_e$		& 0.73	& 1.04 	& 1.30\\
  $PA$ 		& -2.51 	& 0.36 	& 4.00\\
  $q$  		& -0.07 	& -0.01 	& 0.06\\
 \hline
  \textbf{SE, S\'ersic}		&		&		& \\
  $R_e$		& 0.72	& 0.98 	& 1.21\\
  $n$			& 0.63	& 0.90 	& 1.48\\
  $PA$ 		& -8.31 	& 0.04 	& 6.43\\
  $q$  		& -0.07 	& -0.01 	& 0.03\\
 		\hline
	\end{tabular}
\end{table}

\subsection{Flipped galaxies}\label{sec:flipped}
In the MPP-VAC, \citet{Fischer_2019} described the phenomenon of `flipped' galaxies, where the S\'ersic component models the more extended structure while the exponential component was found to be more compact. They identified the flipped galaxies as those having  $R_{e,\text{exp}} < R_{e,\text{S\'ersic}}$ and $n_{\text{S\'ersic}}<1$, and thus labelled the bulge and disc components in their fits accordingly. 

\begin{figure}
 \includegraphics[width=0.7\linewidth]{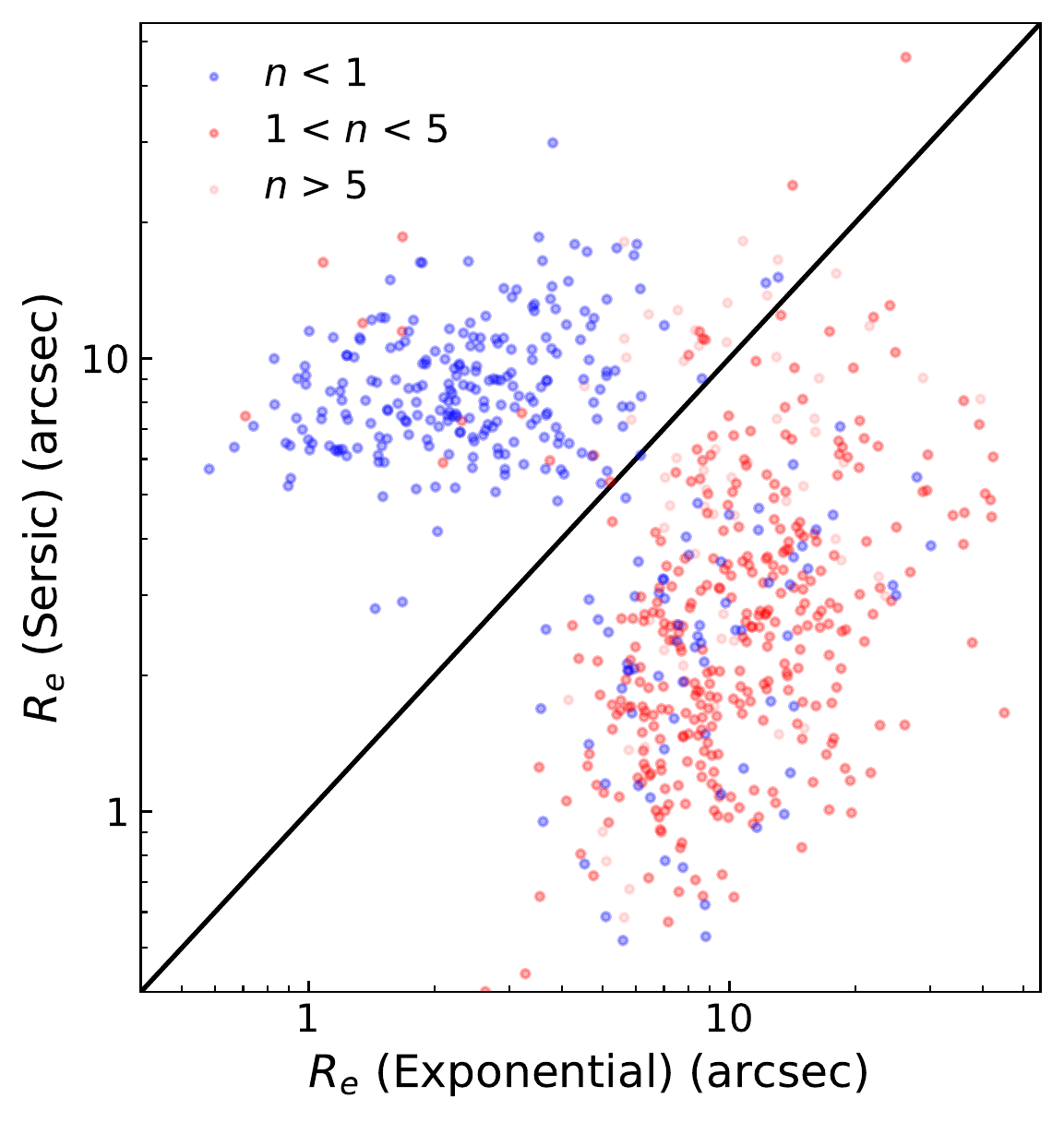}
 \caption{Comparison of the $R_e$ for the S\'ersic and exponential components in each SE fit, with blue, red and light red points representing $n<1$, $1<n<5$ and $n>5$, respectively. }
 \label{fig:BD_Re}
\end{figure}

This phenomenon is explored in this data set in Figure~\ref{fig:BD_Re}, which compares the $R_e$ of the S\'ersic and exponential components, colour-coded according to the S\'ersic index of the S\'ersic model. It can be seen that the majority of galaxies with $R_{e,\text{exp}} < R_{e,\text{S\'ersic}}$ fall into two categories- those with very high S\'ersic indices (i.e. $n>5$), representing bright cores and broad wings ($\sim20\%$ of the sample, displayed as light red points), and those with S\'ersic indices $n<1$ ($\sim75\%$ of the sample, blue points). Consequently, in the majority of cases where the S\'ersic component models the more extended structure within the galaxy, it also has a S\'ersic index of below 1. Clearly, this scenario does not fit with the assumption that the S\'ersic and exponential components always model the bulge and disc respectively, but is in agreement with the findings of \citet{Fischer_2019}.

\begin{figure}
 \includegraphics[width=1\linewidth]{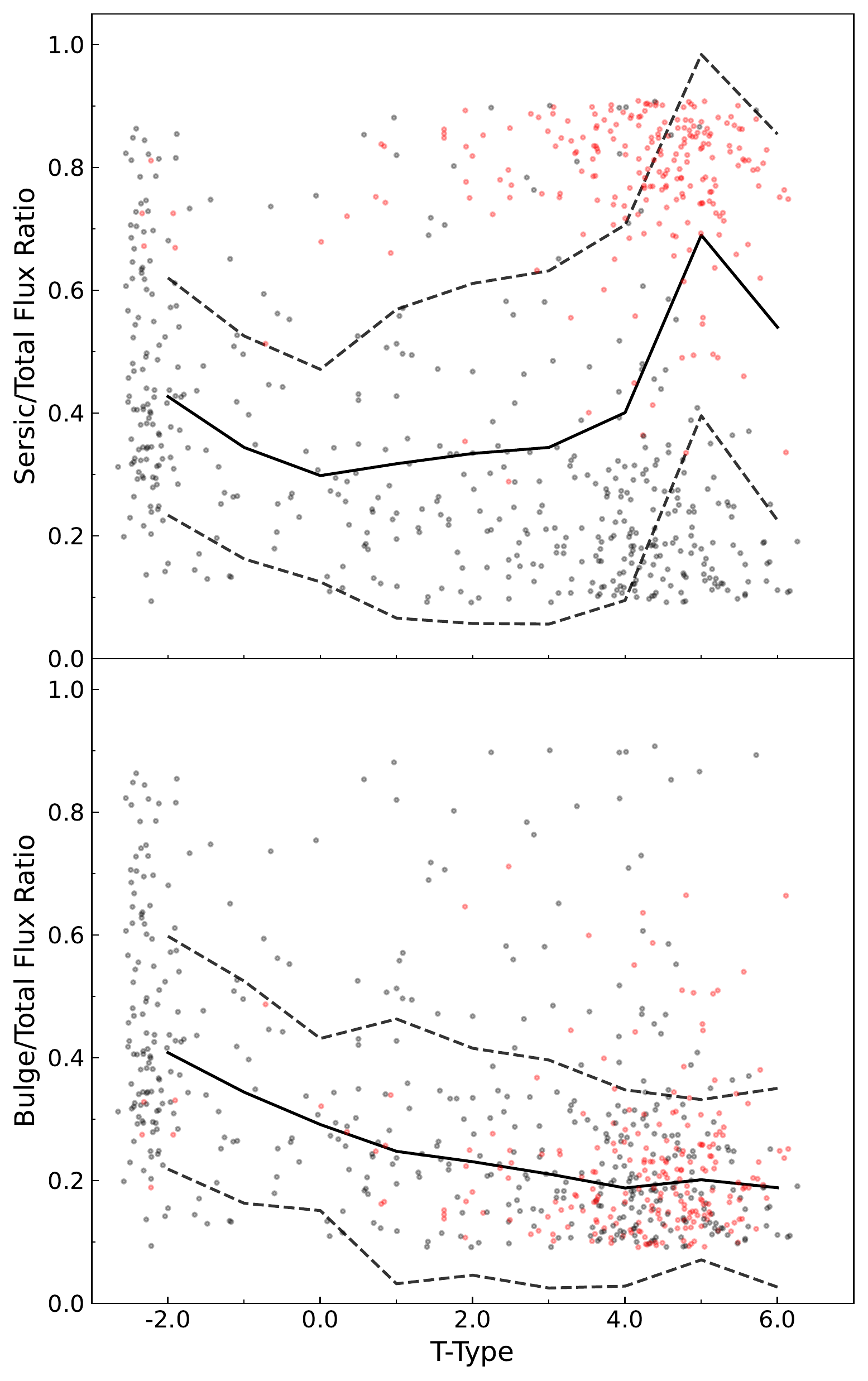}
 \caption{The S\'ersic-to-Total (top) and Bulge-to-Total (bottom) flux ratios of each galaxy as a function of T-Type. The solid line represents the rolling median, with the dashed lines showing $\pm1\sigma$. The red points mark the galaxies that were identified as being flipped.}
 \label{fig:BT_morph}
\end{figure}

We followed \citet{Fischer_2019} and identified flipped galaxies as those with $n_{\text{S\'ersic}}<1$ and $R_{e,\text{exp}} < R_{e,\text{S\'ersic}}$. However, before officially identifying these galaxies as flipped, a final consistency check was carried out. In Fig.~\ref{fig:BT_morph}, the S\'ersic-to-Total and Bulge-to-Total flux ratios for each galaxy are plotted against the morphology. In this plot, the flipped galaxies are highlighted in red. If the bulges and discs have been correctly identified, one would expect that the Bulge-to-Total flux ratio would decrease with increasing T-Type as one moves from early to late type galaxies. This trend can clearly be seen in the plot for the Bulge-to-Total flux ratios, while the S\'ersic-to-Total light ratios are actually seen to increase for the later type spirals. Consequently it can be seen that in this sample, the identification of flipped galaxies appears to work, and that the majority of the flipped galaxies have later type morphologies.

In this sample,  $\sim33\%$ of the galaxies are considered to be flipped. This number is significantly higher than the $\sim13\%$ found by \citet{Fischer_2019}, which may be due to a combination of factors. For example, the MaNGA images used for the fits in this work use a smaller sample of galaxies with a smaller FOV and lower spatial resolution, which may introduce some inherent biases to the fits. Furthermore, it could also be due to differences in the fitting procedures, where \citet{Fischer_2019} carried out repeated fits for galaxies with $n_{\text{S\'ersic}}=8$ to constrain the upper limit for the S\'ersic index. This additional step was not carried out in this study, and may have some effect on whether some galaxies are flipped. And finally, the initial parameters used for the fits with \textsc{buddi} came from the MPP-VAC, and so if a galaxy was flipped in that sample, the \textsc{buddi} fit also started as being flipped.

Due to the presence of likely flipped galaxies in both this work and the MPP-VAC, we follow the methodology of  \citet{Fischer_2019} to report the sizes of both components in terms of $R_e$ instead of using the scale length, $R_d$, for the size of the exponential component. However, should one want to use the scale length of the exponential component, it can be calculated from the effective radius using $R_d = R_e/1.678$ \citep{Meert_2015}.

\section{Overview of the Stellar Population Properties}\label{sec:overview_stellar_pops}
Having cleanly extracted the spectra for each component and identified the galaxies with good fits, the next step was to derive the stellar populations. A detailed analysis of the stellar populations in each component as a function of galaxy morphology, mass etc is beyond the scope of this work, and will be explored in future papers. However, an overview of the basic stellar populations properties is presented here, alongside some further tests of the reliability of the fits using a more statistical sample than was available in \citet{Johnston_2017}.

\subsection{Analysis of the Mass-Weighted Stellar Populations}\label{sec:analysis_stellar_pops}

The mass-weighted stellar populations were measured from each spectrum through full-spectral fitting using \textsc{ppxf}. A regularized fit was applied to each spectrum using a linear combination of template spectra of known relative ages and metallicities from the MILES evolutionary synthesis models of \citet{Vazdekis_2015}. These model spectra span an age and metallicity range of $0.03-14\,$Gyr and  [M/H]=$-2.27$ to +0.40 respectively, and were created using the BaSTI isochrones \citep{Pietrinferni_2004,Pietrinferni_2006}.  
The fits were also carried out with the template spectra created using the Padova isochrones \citep{Girardi_2000}, but while the results were consistent between the two sets of models, the sparser sampling in the high-metallicity regime led to some artefacts in the metallicities measured after regularization. Consequently, the results presented in this section were derived using the BaSTI isochrone templates. The best fit models were convolved with a multiplicative Legendre polynomial of order 10 to model the shape of the continuum, thus reducing the sensitivity to dust reddening and omitting the requirement of a reddening curve \citep{Cappellari_2017}.

The regularization process is designed to smooth the variation in the weights of the templates with similar ages and metallicities. This step reduces the chances of template mismatch, which contributes to the noise in the stellar population parameters, and attaches a physical meaning to the weights assigned to the templates in term of the star formation history, age and metallicity distribution within an individual galaxy. In order to determine the level of regularization required, an unregularized fit is first applied to each spectrum to measure the $\chi^2$ value for the fit. The noise spectrum was then scaled appropriately until $\chi^2/N_{DOF} = 1$, where $N_{DOF}$ is the number of degrees of freedom in the fit (i.e. the number of unmasked pixels in the input spectrum). Finally, the fit to the spectrum was repeated using the scaled noise spectrum and increasing the regularization value in steps until the $\chi^2$ of the fit increased by $\Delta\chi^2 = \sqrt{2\times N_{DOF}}$. This value represents the limit between a smooth fit that still reflects the star-formation history of the galaxy and one that has been smoothed excessively. However, it is important to note that this fit may not reflect the true star-formation history of the galaxy, in particular for variations in the star-formation activity over shorter timescales than the models allow. Consequently, these regularized fits are designed to reduce the age-metallicity degeneracy between spectra.
The template spectra used in these fits are modelled for an initial birth cloud  of $1 \text{M}_\odot$, meaning that the final weight of each template reflects the `zero-age' mass-to-light ratio of that stellar population. As a result, the smoothed weighting of  template spectra with similar ages and metallicities represents a simplified star-formation history of the component being modelled by defining the relative mass contribution of each stellar population \citep[see e.g.][]{McDermid_2015}.

\begin{figure*}
 \includegraphics[width=0.9\linewidth]{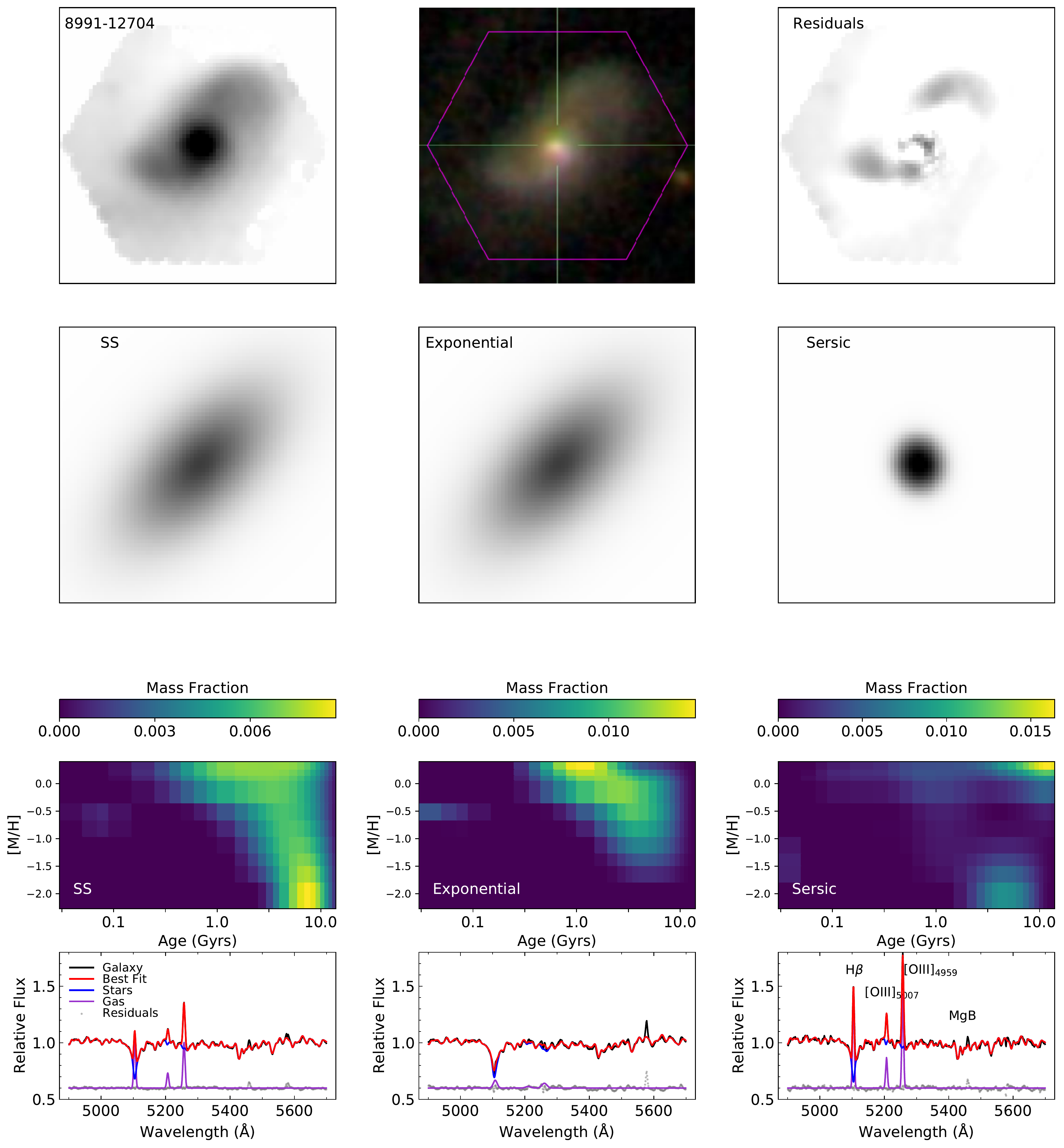}
 \caption{An example of the fit to galaxy 8991-12704. The top row shows the MaNGA white-light image (left), and SDSS $gri$-band image with the MaNGA FOV outlined (middle), and the white-light residual image from the SE fit (right). The second row shows the best fit for the SS model (left) and the exponential and S\'ersic components for the SE fit (middle and right). For each model, the smoothed age-metallicity grid and the \textsc{ppxf} fit to the integrated spectrum for that component are given below in the bottom two rows. Note that in the bottom row, the gas spectrum and the residuals have been offset for clarity, and several key emission and absorption lines have been identified in the fit to the S\'ersic component spectrum for reference. 
}
 \label{fig:spec_2}
\end{figure*}

An example of the regularized SFH derived for the spiral galaxy with the MaNGA plate-IFU number 8991-12704  is shown in Fig.~\ref{fig:spec_2}. This figure includes the MaNGA white-light image of the galaxy alongside the SDSS $gri$-band image, the residual image after subtracting the SE model, and the best-fitting model images for the SS model and the exponential and S\'ersic components in the SE fit. Below each model image are the smoothed age-metallicity weights grids, which represent the SFHs for the spectra extracted using the corresponding models, and the fit to the spectrum by \textsc{ppxf}. For this galaxy, one can see that the SFH derived from the SS fit shows that the galaxy has undergone an extended star formation history, with the majority of the mass being built up between $1-10$~billion years ago. When we consider the SFHs of the S\'ersic and exponential components, their independent SFHs become clear. The exponential component shows generally younger stellar populations than the S\'ersic component, and was built up over a longer timescale, while the S\'ersic component shows that the majority of its stellar mass was formed early in the lifetime of the galaxy with only a tiny fraction of the mass formed more recently. These results are consistent with the model that the exponential and S\'ersic components represent the disc and bulge respectively, and the typical scenario where spiral galaxies contain an old bulge surrounded by a star forming disc. Thus, this figure demonstrates the strength of using \textsc{buddi} to cleanly disentangle the stellar populations of distinct components within galaxies.

\begin{figure*}
 \includegraphics[width=0.9\linewidth]{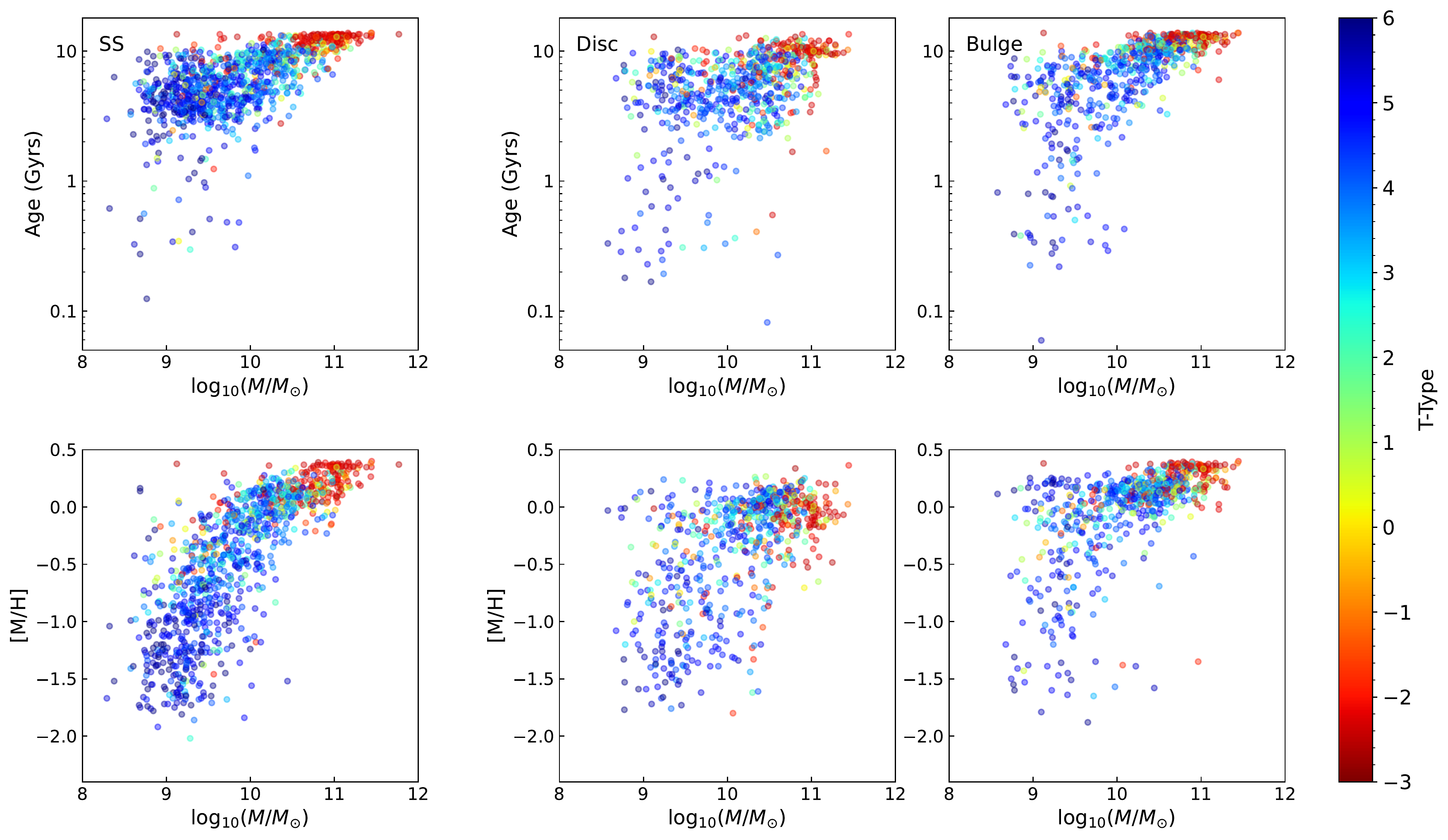}
 \caption{Ages (top) and metallicities (bottom) as a function of the total galaxy mass for the SS fits (left) and the discs and bulges of the SE fits (middle and right respectively), colour-coded by the T-Type of each galaxy.  }
 \label{fig:bd_pops_mass}
\end{figure*}

In order to compare the stellar populations of the bulges and discs across the entire sample, the mean mass-weighted ages and metallicities of each galaxy as a whole (from the SS fit) and  of their bulges and discs (from the SE fit) were calculated using 
\begin{equation} 
	\text{log(Age$_{\text{M-W}}$)}=\frac{\sum \omega_{i} \text{log(Age$_{\text{template},i}$)}}{\sum \omega_{i}}
	\label{eq:age}
\end{equation}
and 
\begin{equation} 
	\text{[M/H]$_{\text{M-W}}$}=\frac{\sum \omega_{i} \text{[M/H]}_{\text{template},i}}{\sum \omega_{i}}
	\label{eq:met}
\end{equation}
respectively, where $\omega_{i}$ represents the weight of the $i^{th}$ template (i.e. the value by which the $i^{th}$ stellar template is multiplied to best fit the galaxy spectrum), and [M/H]$_{\text{template},i}$ and Age$_{\text{template},i}$ are the metallicity and age of the $i^{th}$ template respectively. 

Figure~\ref{fig:bd_pops_mass} presents an example of the mass-weighted ages and metallicities for all galaxies successfully fitted with the SS and SE models, plotted against the mass of the total galaxy and colour coded according to the morphology in terms of T-Type (red indicates earlier types while blue indicates later type morphologies). In the plots for the SS fits, one can clearly see the red and blue sequences, such that the higher-mass ETGs generally have older, more metal-rich stellar populations while the lower-mass LTGs have younger, more metal-poor stellar populations. Furthermore, the distributions in the ages and metallicities of the LTGs are much broader than the ETGs. These trends are in agreement with the stellar mass-metallicity and mass-age relations in the literature. For example, the first study of the stellar mass-metallicity relation was by \citet{Gallazzi_2005}, in which they found that low-mass galaxies tended to be younger and more metal-poor than the higher-mass galaxies, with a rapid transition between these two regimes occurring for the mass range $3\times10^{9}-3\times10^{10}$M$_\odot$. This same transition in age and metallicity can be seen in the SS plots in Fig.~\ref{fig:bd_pops_mass}. A recent study by \citet{Zenocratti_2022} of the mass-metallicity relation using galaxies in the EAGLE simulation also found a link with morphology, such that at a given mass, the rotationally-supported disc galaxies tend to have lower metallicities than the dispersion supported spheroidal galaxies. They concluded that this trend arose due to the disc galaxies having recently accreted low-metallicity gas, which fuelled more recent star formation, while the spheroidal galaxies instead formed stars earlier on by using their own gas reservoirs.

When looking at the plots for the SE fits, the same general trends appear to also hold true for the bulges and discs, though with more scatter. For example, in the higher mass galaxies ($M>10^{10}M_\odot$), the discs show a broader distribution in both age and metallicity than the bulges. This trend could indicate fundamental differences in the evolution of the bulges and discs themselves in this mass range, potentially reflecting  more recent star formation in the discs of high-mass galaxies fuelled by accreted low-metallicity gas while the bulges have evolved more passively, as was already proposed by \citet{Zenocratti_2022} for disc-dominated galaxies. In low-mass galaxies however, both the bulges and discs show evidence of younger and more metal-poor stellar populations, indicating that both components may have formed more recently than in the higher mass galaxies. A more detailed analysis of the bulge and disc star-formation histories in the S0s in this sample by Johnston~et al (submitted) have indeed found this trend, where the bulges of galaxies with masses M~$>10^{10}$M$_\odot$ formed rapidly in the early lifetime of the galaxy, while those in lower mass galaxies formed more recently over longer timescales. Additionally, a more detailed analysis of this trend for spiral galaxies is ongoing, and will be presented in Jegatheesan~et al (in prep). However, similar trends have been seen in studies that applied bulge-disc decomposition techniques to photometric data. For example, a recent study by \citet{Robotham_2022} used the \textsc{ProFUSE} package to explore the mass-size-age plane for galaxy components through SED fitting to $ugri$ZYJHKs imaging from the GAMA survey, finding a strong trend in the disc metallicity with the disc mass such that higher mass discs have higher metallicities, as seen in Fig.~\ref{fig:bd_pops_mass}.


\subsection{Investigating the effect of the Chebychev polynomials}\label{sec:poly_effect}
In Section~\ref{sec:step_3} we discussed the Chebychev polynomials selected for the fit parameters, specifically using a polynomial of order 1 for the $R_e$ and $n$ in the SE fits and an order of 2 for these parameters in the SS fits (the $q$ and $PA$ used polynomials of order 1 throughout). This additional degree of freedom was selected for the SS fits to improve the models of 2-component galaxies by allowing for colour gradients. Such gradients could be created, for example, in spiral galaxies with old, red bulges surrounded by younger, star-forming discs, making them appear larger at bluer wavelengths and more compact in the redder end of the spectrum. Since these polynomials model the general shape of the light distribution throughout the galaxy, the choice of polynomial between orders 1 and 2 should not affect the strength of the spectral features significantly. 

\begin{figure}
 \includegraphics[width=\linewidth]{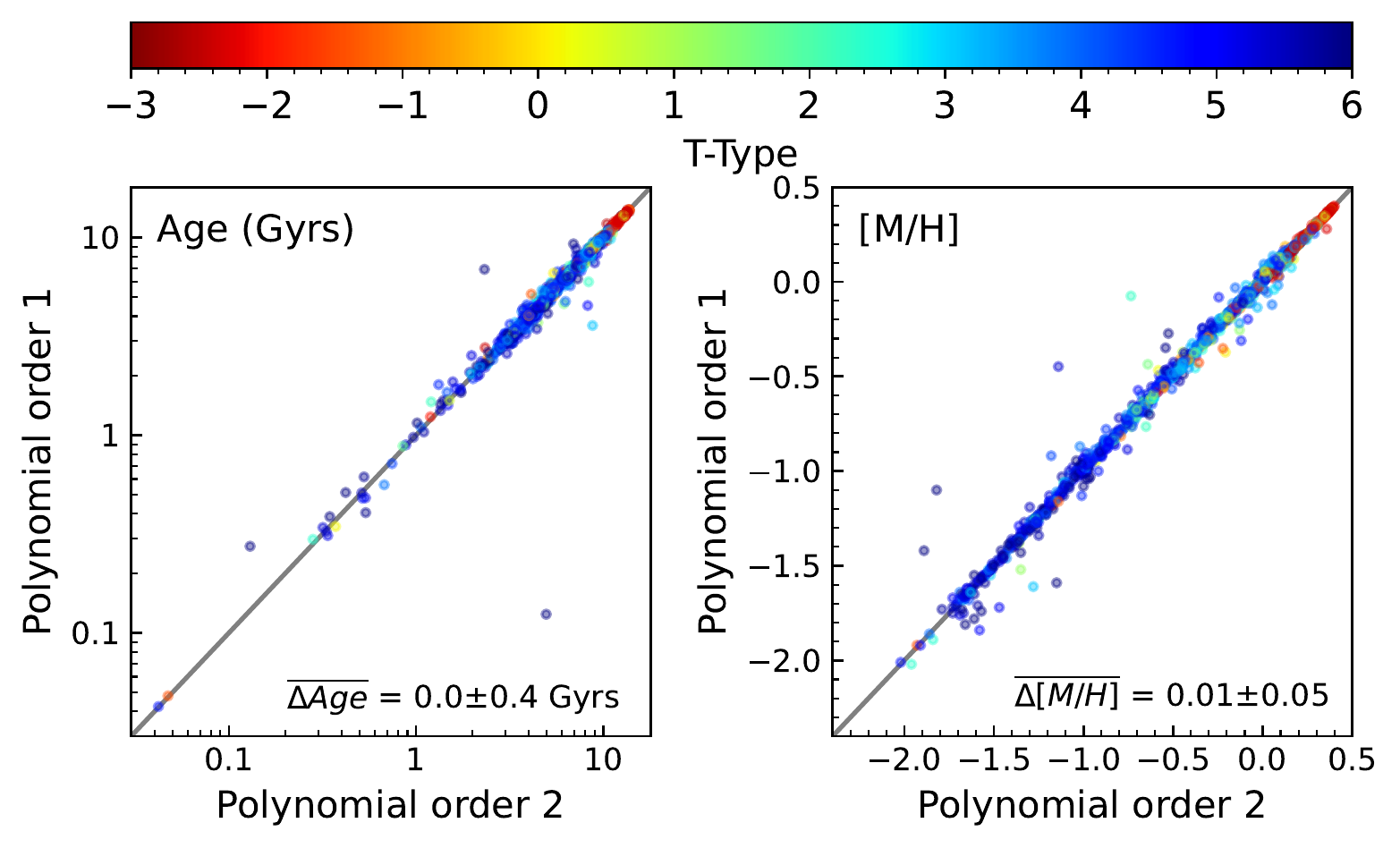}
 \caption{A comparison of the mass-weighted ages (left) and metallicities (right) for the spectra derived from the SS fits using Chebychev polynomials of orders 1 and 2 to model the variation in $R_e$ and $n$ with wavelength. The mean differences in the age and metallicity measurements are given in the bottom right of each plot, with the uncertainty reflecting the standard deviation of the scatter.}
 \label{fig:poly_comp}
\end{figure}

In order to test this hypothesis, the SS fits were repeated using a polynomial of order 1 for the $R_e$ and $n$, and the stellar populations derived in the same way. Apart from the choice of polynomial, the fits and the analysis were identical. Figure~\ref{fig:poly_comp} compares the ages and metallicities of the galaxies using these two polynomial orders, and show that the results are generally consistent, with standard deviations of 0.38~Gyrs in age and 0.05 in metallicity. Consequently, we can say that the stellar populations derived from the spectra extracted by \textsc{buddi} are fairly insensitive to the choice of polynomial between an order of 1 and 2.

\section{Comparison with MPP-VAC}\label{sec:BUD_PYM_comparison}
\begin{figure*}
 \includegraphics[width=0.9\linewidth]{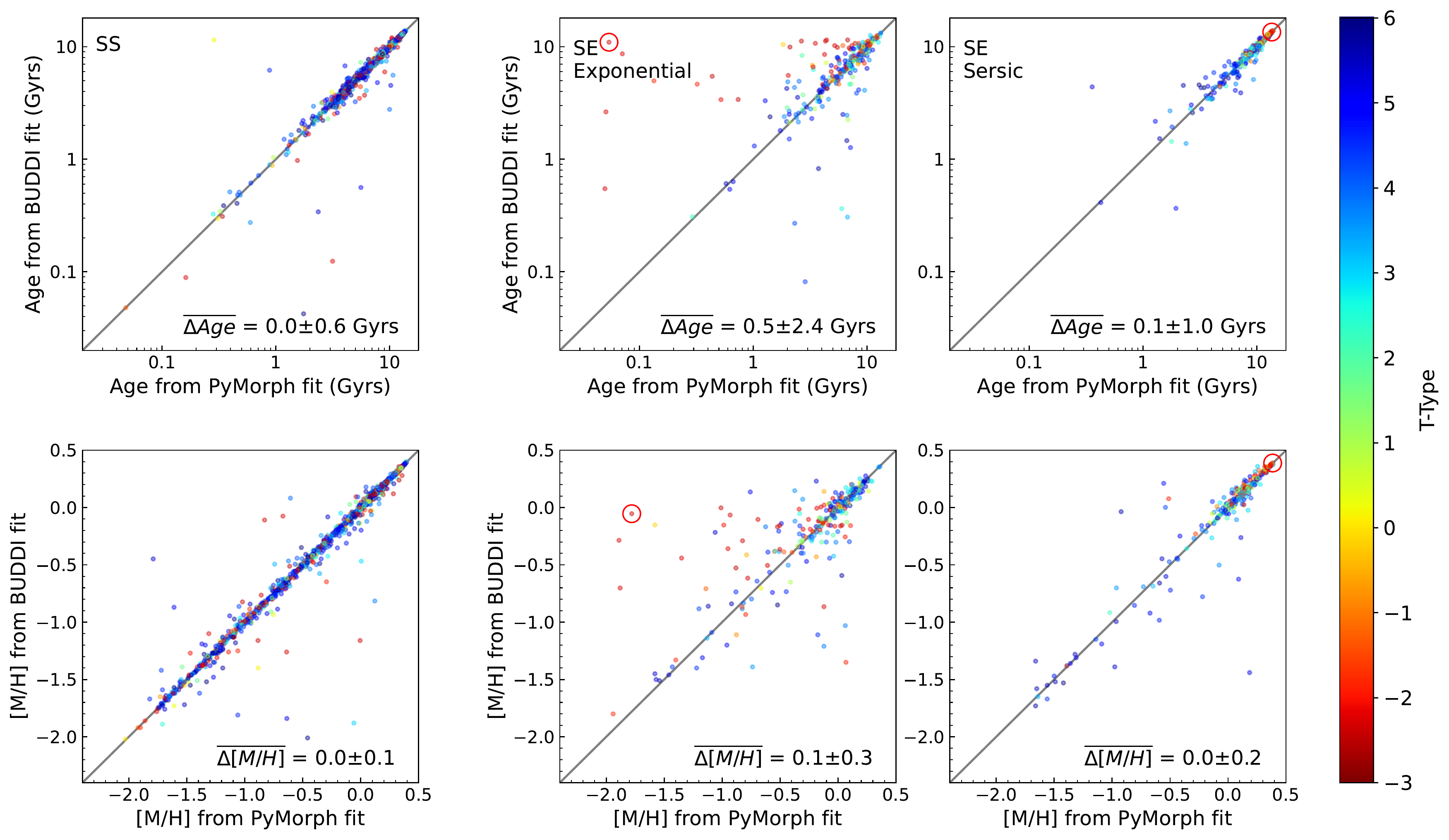}
 \caption{Comparison of the stellar populations for each component extracted using the \textsc{buddi} and \textsc{pymorph} fit parameters. The colours correspond to the morphology, as indicated in the colour bar on the right, with redder colours reflecting ETGs (T-Type$<0$). The mean differences in age and metallicity measurements are given in the bottom right of each plot, with the uncertainty reflecting the standard deviation. The data points with the circle around them identify the measurements for 8249-12705, for which the spectra extracted for the exponential component are plotted in Fig.~\ref{fig:buddi_pym_comp2}. 
 }
 \label{fig:buddi_pym_comp}
\end{figure*}

In section~\ref{sec:overview} we explored the differences in the structural parameters between the fits obtained by \textsc{buddi} and those in the MPP-VAC that were derived from SDSS images of the MaNGA galaxies using the \textsc{pymorph} code. While no significant differences were detected in the trends for the structural parameters, even small differences in the fits could potentially lead to significantly different estimates of the stellar populations in each component. While this issue was explored in \citet{Johnston_2017} for the prototype MaNGA data, the small number of galaxies available at the time meant that the tests were carried out on a series of simulated datacubes. In this section, we will explore this issue in more detail by comparing the stellar populations derived from the spectra extracted using the \textsc{buddi} and MPP-VAC fit parameters.

For this test, the galaxies were modelled again with \textsc{buddi} using the fit parameters for each component in the MPP-VAC, and with 0-order polynomials for the $R_e$, $n$, $q$ and $PA$ (i.e. holding them fixed at the input values along the entire wavelength range) and allowing the magnitudes complete freedom. The mean mass-weighted ages and metallicites of the resulting bulge and disc spectra were derived in the same way as described in Section~\ref{sec:overview_stellar_pops}, and are plotted in Fig.~\ref{fig:buddi_pym_comp} for both the SS and SE fits. As before, the colours reflect the morphology of each galaxy. It should be noted that these plots only include the results for spectra that were derived from successful fits in both cases, using the criteria outlined in Section~\ref{sec:success}, where the parameters derived by \textsc{buddi} and the MPP-VAC were assessed independently. The mean differences between the ages and metallicities measured in the two sets of fits, along with the standard deviation of the scatter, are given in the bottom right of each plot.

One can see that, in general, the measurements for the ages and metallicities for the SS and SE fits are very consistent, though the SE fits do show a larger scatter, particularly for the exponential component.  The colours of the data points show that many of the outliers in the results for the exponential component are early-type galaxies, with T-Types$<0$, and so it is possible that in some cases these poor results are due to forcing a 2-component fit on an elliptical galaxy that is already well modelled with a single S\'ersic profile, which makes profiles fits inherently unreliable. The outliers in the S\'ersic components of the SE fits, on the other hand, tend to be LTGs, but the correlation is much tighter in this case. Since Fig.~\ref{fig:BT_morph} shows that the majority of flipped galaxies have late type morphologies, it is possible that this scatter is due to mismatched components, i.e. where the S\'ersic and exponential components have been flipped in one set of fits but not the other. However, only $\sim4\%$ of these galaxies  were found to fall into this category (i.e. flipped in one set of fits but not the other), and so this mismatch does not completely explain the scatter.

\begin{figure*}
 \includegraphics[width=\linewidth]{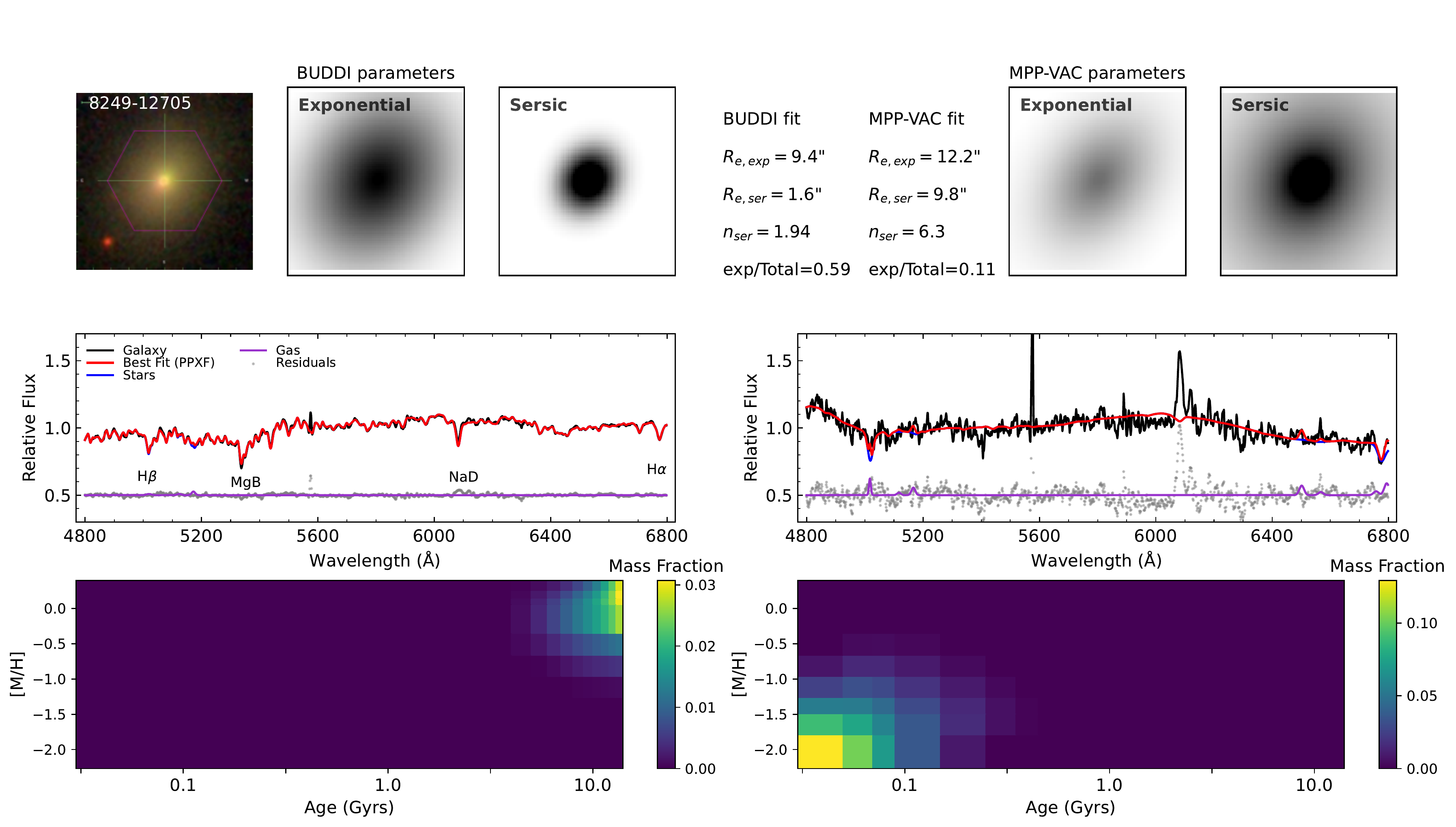}
  \caption{An example of the fits to galaxy 8249-12705 using the parameters derived using \textsc{buddi} (left) and with those from the MPP-VAC (right). The top row shows the SDSS $gri$-band image of galaxy alongside the model images for the exponential and S\'ersic components from each fit. The key fit parameters for the two fits ($R_e$, $n_{S\'ersic}$ and exponential-to-total light ratio) are also given. The middle row shows the spectrum extracted for the exponential component in the SE fit using both sets of fit parameters, along with the best fit from \textsc{ppxf} superimposed in red. The bottom row shows the corresponding smoothed age-metallicity grid for the fits to the spectra in the middle row.}
 \label{fig:buddi_pym_comp2}
\end{figure*}

To  further explore the outliers, Fig.~\ref{fig:buddi_pym_comp2} has been replotted in Figures~\ref{fig:buddi_pym_comp_n} and  \ref{fig:buddi_pym_comp_n} in Appendix~\ref{sec:Appendix_A}, this time colour coding the points according to the ratios of the values derived by \textsc{buddi} to those from the MPP-VAC. While no obvious trend can be seen between the difference in the stellar populations and these fit parameters, it is interesting to note that the outliers in the exponential components that are marked as ETGs in Fig.~\ref{fig:buddi_pym_comp2} also appear to show lower ratios in the $R_e$ and $n$ values between the two sets of fits, i.e. the values derived by \textsc{buddi} are significantly lower than those in the MPP-VAC. Again, this effect may be a result of forcing a 2-component fit upon galaxies that are better modelled with a single S\'ersic component.

A closer inspection was carried out on the fits to some of the outlying points to better understand the discrepancy in their measurements. The largest outlier in the age and metallicity plots for the exponential components corresponds to galaxy 8249-12705 (marked with a circle around the point in Fig.~\ref{fig:buddi_pym_comp}), and the corresponding spectra for the exponential component derived using the  \textsc{buddi} and MPP-VAC fit parameters are shown in Fig.~\ref{fig:buddi_pym_comp2}. This figure also shows the SDSS $gri$-band image of this galaxy and the model images for each component from both sets of fit parameters. It can be seen that despite using the MPP-VAC parameters as the initial estimates, \textsc{buddi} has converged on a very different fit where the S\'ersic component is much more compact and with a lower S\'ersic index, and where the exponential component is brighter (the exponential-to-total light ratio is 0.59 for the \textsc{buddi} fit compared to 0.11 for the MPP-VAC fit). Below these images, the spectra derived for the exponential component are given, with the best fit by \textsc{ppxf} over plotted in red, and the smoothed age-metallicity grid is also given, showing the weights of the template spectra used in that fit.  It can be seen that the spectrum extracted by \textsc{buddi} has higher S/N with clear spectral features, whereas the  spectrum created using the MPP-VAC fit parameters is much noisier with few clear spectral features. This lower S/N for the MPP-VAC fit is simply due to the faintness of the exponential component in the model, where the S\'ersic component dominates the light in all spaxels in the FOV, and which is not the case for the \textsc{buddi} fit. As a result, the fit to the MPP-VAC exponential spectrum is very poor, and produces significantly younger, more metal-poor measurements for the stellar populations. The morphology of this galaxy is listed in the MDLM-VAC as T-Type$=-2.4$ and a probability of being an S0 of 0.2, making it a likely Elliptical galaxy. As a result, any 2-component fits to this galaxy should be used with care and should not be assumed to represent the bulge and disc components. 
In general, though, where good fits are achieved with the structural parameters derived by \textsc{buddi} and using those from the MPP-VAC, the results, in terms of both the fit parameters and the stellar populations derived from the spectra, are generally consistent. Consequently, we conclude that the \textsc{buddi} fits are reliable, and that the stellar populations derived truly reflect those of the components modelled in the fits. These results also show that \textsc{buddi} can be used for IFU data with less coverage of the galaxy if the structural parameters can be measured reliably from broad-band photometric data and held fixed within \textsc{buddi}.

A natural question occurs here- why use \textsc{buddi} for the full fits when it's been shown that you can achieve similar results using fit parameters from catalogs such as the MPP-VAC, or from applying your own fits to imaging data of the same galaxies? The main advantage of using \textsc{buddi} is that it is self contained- imaging data of the targets may not be available over the same wavelength range or to the same depth as the IFU datacube. By carrying out the fits within \textsc{buddi}, you can ensure that you have derived reliable fit parameters over the wavelength range you are actually using, and avoid introducing errors or bias when converting pixel scales etc between the imaging and spectroscopic data. Fits to simulated data by \citet{Haeussler_2022} have shown that such multi-band fitting significantly improves the measurements of the parameter values, particularly of the effective radii and S\'ersic indices, when compared to fitting a single image. Furthermore, running the initial fits within \textsc{buddi} also allows the user more flexibility to modify the initial parameters, include or exclude components, and to update the polynomials for the fit parameters as a function of wavelength to ensure the best fit to the datacube for their science case. While this step is also possible with separate imaging data, differences in the depth of the images and the spatial resolution may result in poor fits when one tries to include too many components in the fit to the IFU data. 

Of course, using \textsc{buddi} in this way assumes you are using IFU datacubes with sufficient spatial resolution and a wide field-of-view. \textsc{buddi} has so far only been tested on data from MUSE and the  MaNGA and CALIFA surveys. It is likely that additional imaging data is required for the  fits with \textsc{buddi} to datacubes from IFU instruments with smaller fields of view, such as SAURON \citep{Bacon_2001}, KCWI \citep{Morrissey_2018} and the SAMI survey. \textsc{buddi} has therefore been written to allow the user the flexibility to define the structural parameters for the fits based on alternative imaging data for such scenarios.


\section{Discussion and Conclusions}\label{sec:conclusions}
We have  used \textsc{buddi} to carry out  automated light profile fits to IFU datacubes of a large sample of galaxies observed as part of the SDSS-IV MaNGA Survey and released as part of the SDSS DR15. We used a single S\'ersic (SS) and S\'ersic+exponential (SE) models to cleanly extract the spectra of the galaxy and the bulge and disc respectively. As the first paper in this series, the purpose of this work is to introduce the BUDDI-MaNGA project, focussing on the procedure for the fit, characterising the final sample, and testing the results. Future papers will focus more on the scientific results of this data set. 

We used the publicly available MaNGA data from the DR15, initially selecting only those galaxies observed with the 91 and 127-fibre IFUs and which had successful fits in the MPP-VAC. The fits with \textsc{buddi} were carried out using the MPP-VAC parameters as initial estimates for the structural parameters. After modelling all the galaxies, we identified a final sample of galaxies with `good' fits, using criteria based on physically meaningful structural parameters. This final sample  consisted of 1038 galaxies for the SS model and 691 galaxies for the SE model. The spectra extracted using the SS model represent the mean global spectrum of the galaxy, while those created using the SE model represent the bulge  and disc components.

The first step in the analysis was to carry out a comparison of the fit parameters derived by \textsc{buddi} with the initial parameters from the MPP-VAC. We found similar trends in their distributions, indicating that \textsc{buddi} was able to model the light profiles of the galaxies reliably despite the smaller FOV of the MaNGA data. The main  difference was that the \textsc{buddi} fits resulted in more S\'ersic components with $n\sim8$ while the MPP-VAC showed a higher frequency of $n<2$. This trend is mainly down to variations in the approach towards the fits between these two works-- in this work the galaxies were modelled once, whereas in \citet{Fischer_2019} the fits with high S\'ersic indices were repeated with reduced upper limits on this parameter and then manually checked one by one to determine the best fit. 

We also observed the phenomenon of flipped galaxies, already reported by \citet{Fischer_2019} and \citet{Lange_2016}, where the S\'ersic profile models the more extended component while the exponential profile fits the more compact component. As in \citet{Fischer_2019}, we found that in the majority of cases where the S\'ersic component was more extended than the exponential component, it also had a low S\'ersic index ($n<1$). Consequently, for the remainder of this work, we identified the discs as the more extended component and bulges as the more compact component. In most cases, the bulges and discs corresponded to the S\'ersic and exponential components respectively, except in the cases where $R_{e, \text{exp}} > R_{e, \text{S\'ersic}}$ and $n_\text{S\'ersic} < 1$ where the bulges and discs were taken to be the exponential and S\'ersic components respectively.

Having extracted the spectra for each component, we used \textsc{ppxf} to carry out regularized fits to the spectra in order to derive estimates of their mean mass-weighted stellar populations. While these populations will be analysed in more detail in future work, the mass-metallicity and mass-age plots for the SS fits, representing each galaxy a whole, and the bulge and disc components in the SE fits were presented. The plots for the SS fits showed clear trends such that high mass  and early type galaxies generally contain older and more metal-rich stellar populations, while lower-mass and later-type galaxies are younger and more metal-poor. Since the mass-weighted stellar populations provide information on when the majority of the stellar mass was created in these galaxies, this trend reflects the known trend that many of the ETGs were formed a long time ago, while the LTGs have undergone significant star formation throughout their lives. Similar trends were seen for the two components in the SE fits, but with a larger scatter in particular in the discs of higher mass galaxies. These results reflect that the discs in most galaxies in this sample have built their mass over a longer period than the bulges, likely through a combination of ongoing star formation, accretion and minor mergers.

Two sets of tests were used to investigate the reliability of the fits. In the first case, the SS fits were repeated using Chebychev polynomials of order 1 for the S\'ersic index and effective radius. The stellar populations were derived in the same way, and the mean mass-weighted ages and metallicities compared. Both methods were found to give consistent values, indicating that the spectra extracted by \textsc{buddi} are fairly insensitive to the choice of polynomial between orders of 1 and 2.

The second reliability test focussed on the fit structural parameters themselves, and how they compare to the fits using the values in the MPP-VAC. The SS and SE fits were repeated using the MPP-VAC structural parameters, holding these values fixed and only allowing the magnitude to vary in each case. The stellar populations were derived in the same way and compared with the original fits, and it was found that for galaxies that were considered to have `good' fits with both sets of parameters, the stellar populations were generally consistent. As a result, we encourage the use of \textsc{buddi} to determine the best fit parameters for each galaxy, but acknowledge that in cases where the FOV is small and the coverage of the galaxy within the IFU datacube is poor, using fit parameters derived from photometric data will suffice.

In conclusion, in this paper we present the BUDDI-MaNGA project, in which $\sim1000$ MaNGA galaxies have been modelled successfully with \textsc{buddi} using single S\'ersic and S\'ersic+exponential models to cleanly extract the spectra of each galaxy and their bulges and disc components respectively. To date, this work provides the largest sample of clean bulge and disc spectra extracted from IFU datacubes using the galaxies light profile information. The spectra derived using these parameters and the subsequent stellar populations analysis will be used in future papers in this series to explore the themes of galaxy formation, evolution, and morphological transformations through their effects on the different components within the galaxies.

\section*{Acknowledgements}
We would like to thank the referee for their useful comments which have helped to improve this paper.
E.J.J. acknowledges support from FONDECYT Iniciaci\'on en investigaci\'on 2020 Project 11200263. K.J. acknowledges financial support from ANID Doctorado Nacional 2021 project number 21211770. We gratefully thank the bomberos de Santiago for extinguishing the fire at PUC before it reached the server room where the BUDDI-MANGA data are stored.

Funding for the Sloan Digital Sky Survey IV has been provided by the Alfred P. Sloan Foundation, the U.S. 
Department of Energy Office of Science, and the Participating Institutions. 

SDSS-IV acknowledges support and resources from the Center for High Performance Computing  at the University of Utah. The SDSS website is www.sdss.org.

SDSS-IV is managed by the Astrophysical Research Consortium for the Participating Institutions of the SDSS Collaboration including the Brazilian Participation Group, the Carnegie Institution for Science, Carnegie Mellon University, Center for Astrophysics | Harvard \& Smithsonian, the Chilean Participation Group, the French Participation Group, Instituto de Astrof\'isica de Canarias, The Johns Hopkins University, Kavli Institute for the Physics and Mathematics of the Universe (IPMU) / University of Tokyo, the Korean Participation Group, Lawrence Berkeley National Laboratory, Leibniz Institut f\"ur Astrophysik Potsdam (AIP),  Max-Planck-Institut f\"ur Astronomie (MPIA Heidelberg), Max-Planck-Institut f\"ur Astrophysik (MPA Garching), Max-Planck-Institut f\"ur Extraterrestrische Physik (MPE), National Astronomical Observatories of China, New Mexico State University, New York University, University of Notre Dame, Observat\'ario Nacional / MCTI, The Ohio State University, Pennsylvania State University, Shanghai Astronomical Observatory, United Kingdom Participation Group, Universidad Nacional Aut\'onoma de M\'exico, University of Arizona, University of Colorado Boulder, University of Oxford, University of Portsmouth, University of Utah, University of Virginia, University of Washington, University of Wisconsin, Vanderbilt University, and Yale University.


\section*{Data Availability}
The MaNGA data used in this paper is already publicly available as part of SDSS DR15 and the associated Value Added Catalogs. The BUDDI-MaNGA derived data, including the fit parameters and stellar populations etc, will be made public in the future after we have carried out the fits to the DR17 data. However, anyone interested in using the current data are welcome to contact us to discuss collaboration.



\bibliographystyle{mnras}
\bibliography{buddi_manga.bib} 


\appendix

\section{Further exploration of the effect on stellar populations from the fit parameters}\label{sec:Appendix_A}

\begin{figure*}
 \includegraphics[width=0.9\linewidth]{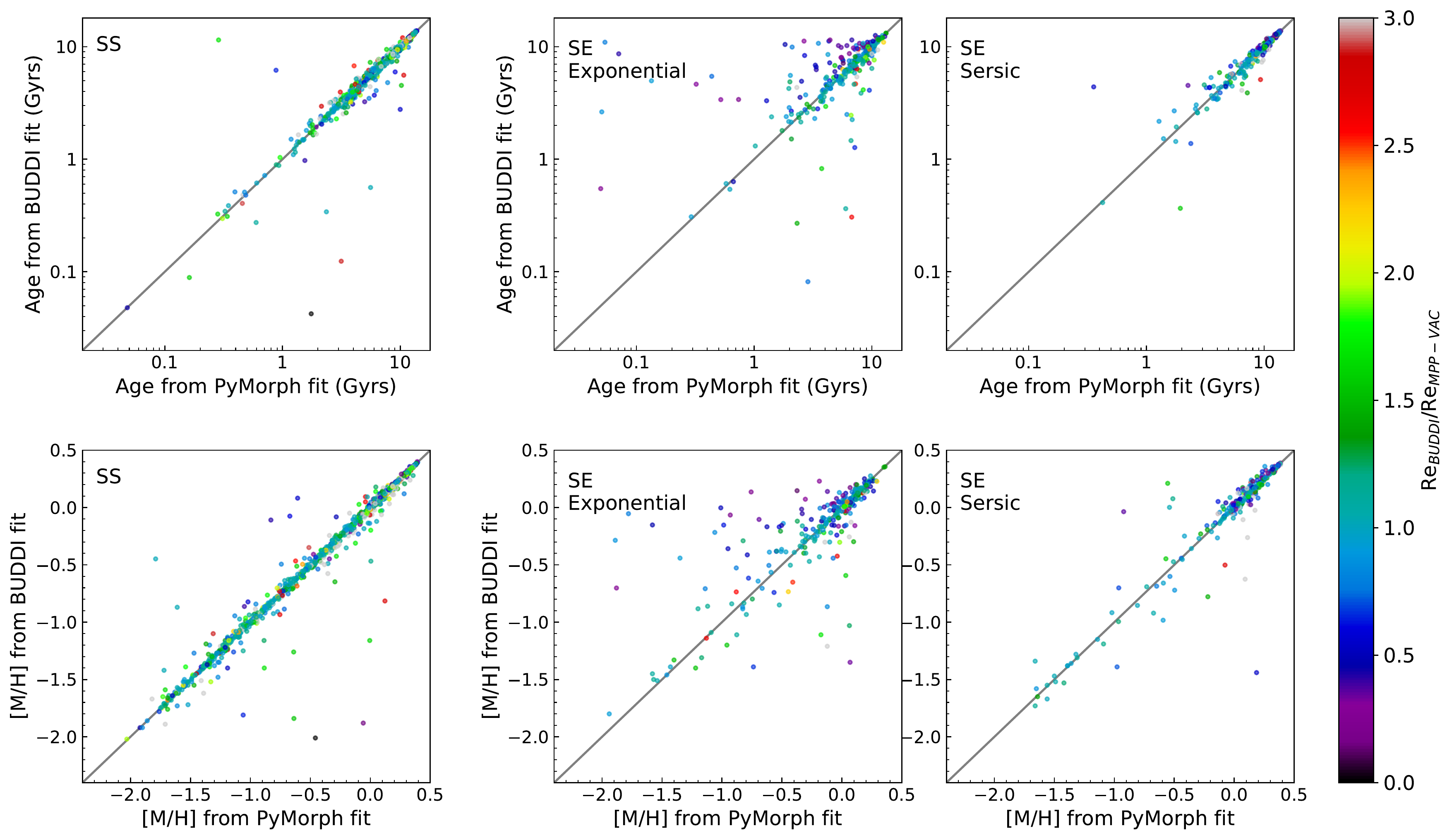}
 \caption{As for Fig.~\ref{fig:buddi_pym_comp}, but colour coded according to the ratio of the $R_e$ values derived by \textsc{buddi} relative to those in the MPP-VAC. 
 }
 \label{fig:buddi_pym_comp_Re}
\end{figure*}

\begin{figure*}
 \includegraphics[width=0.9\linewidth]{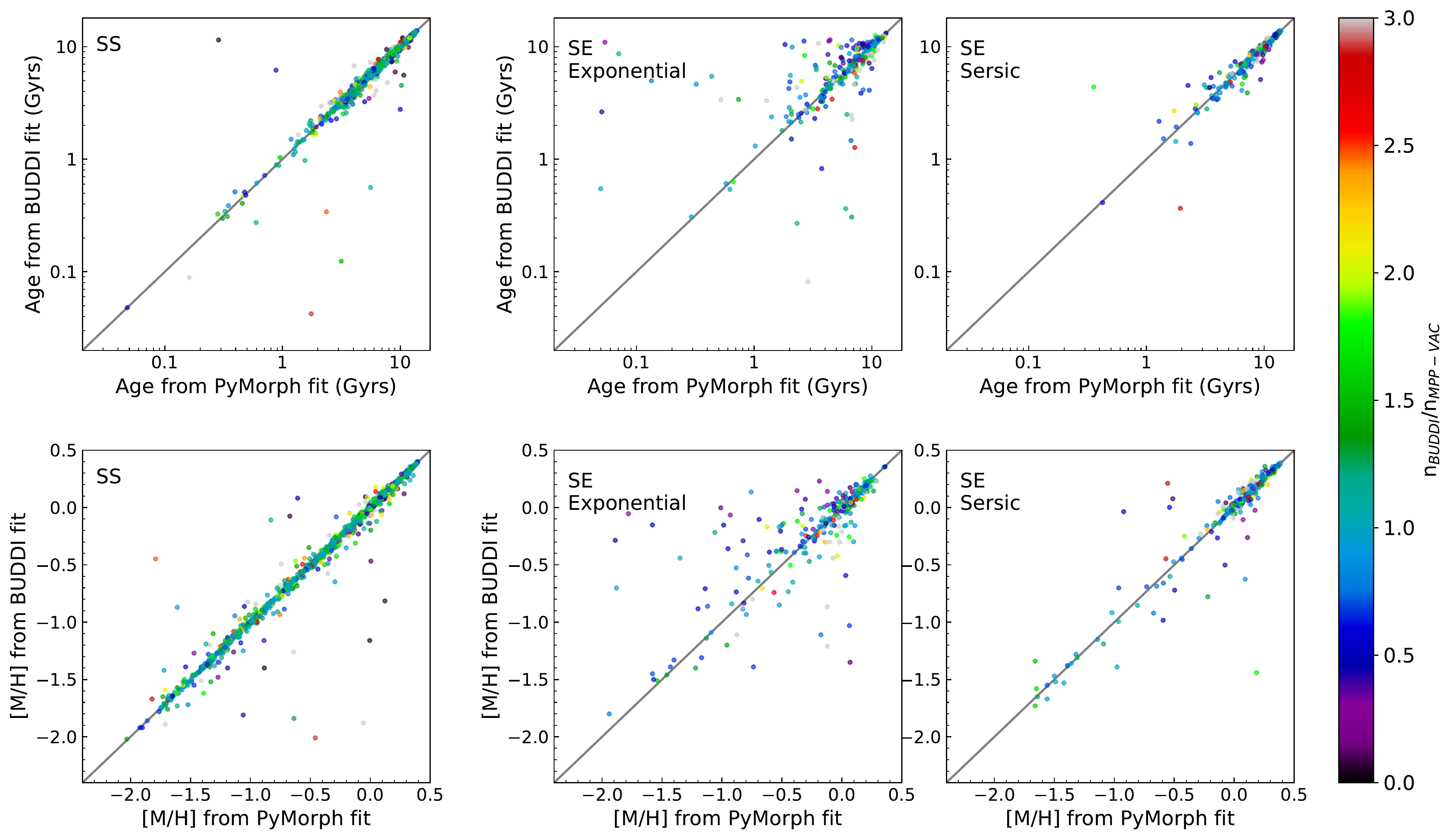}
 \caption{As for Fig.~\ref{fig:buddi_pym_comp}, but colour coded according to the ratio of the $n$ values derived by \textsc{buddi} relative to those in the MPP-VAC. Note that in the case of the exponential component, the ratio of the S\'ersic indices actually corresponds to those of the S\'ersic component.
 }
 \label{fig:buddi_pym_comp_n}
\end{figure*}

In this section, we replot Fig.~\ref{fig:buddi_pym_comp} to compare the stellar populations derived from spectra extracted using the \textsc{buddi} and MPP-VAC fit parameters, now colour coding the points according to the ratio of the $R_e$ (Fig.~\ref{fig:buddi_pym_comp_Re}) and $n$ (Fig.~\ref{fig:buddi_pym_comp_n}) values derived by \textsc{buddi} and the MPP-VAC. Note that the colours for the exponential component in Fig.~\ref{fig:buddi_pym_comp_n} correspond to the ratio in the values for $n$ for the S\'ersic component since this component has a fixed value for $n$.

\bsp	
\label{lastpage}
\end{document}